\documentclass[preprint,preprintnumbers,amsmath,amssymb,floatfix,nofootinbib]{revtex4}
\usepackage{epsf, array, color}
\usepackage{graphicx}
\usepackage{subfigure}
\usepackage{bbold}

\newcommand{\bea}{\begin{eqnarray}}
\newcommand{\eea}{\end{eqnarray}}

\def\beq{\begin{equation}}
\def\eeq{\end{equation}}
\def\call{{\cal{L}}}
\begin{document}
\preprint{CP3-13-49}
\title{Electroweak Effective Operators and Higgs Physics}
\author{Chien-Yi Chen$^{(1)}$, S.~Dawson$^{(1)}$}
\author{Cen Zhang$^{(2)}$}

\affiliation{
$^{(1)}$ Department of Physics, Brookhaven National Laboratory, Upton, New York, 11973
\\
$^{(2)}$ Centre for Cosmology, Particle Physics and Phenomenology (CP3)\\
Universit\'e Catholique de Louvain, B-1348 Louvain-la-Neuve, Belgium
\vspace*{.5in}}

\date{\today}

\begin{abstract}
We derive bounds from oblique parameters on the dimension-$6$ operators of an 
effective field theory of electroweak gauge bosons and the Higgs doublet.  The loop-
induced contributions to the $\Delta S$, $\Delta T$, and $\Delta U$ oblique parameters 
are sensitive to these contributions and we pay particular attention to the role of
renormalization when computing loop corrections in the effective theory.
 Limits on the coefficients of the effective theory from 
loop contributions to oblique parameters yield complementary information to direct
Higgs production measurements.  

\end{abstract}

\maketitle
\section{Introduction}
With the discovery of a particle with the general appearance of a Higgs boson, the focus has turned
to the detailed measurements of the properties of the Higgs boson candidate.
  This involves the measurement of as many Higgs
production and decay channels as possible, along with the limits
inferred from precision electroweak measurements\cite{Baak:2012kk}.  
Currently, all measurements are within $\sim 2-3\sigma$ of
the predictions of the Standard Model (SM).  A deviation from these predictions could be a signal of 
physics beyond the Standard Model.

A search for new physics in the Higgs sector can either involve examination of specific models
or the use of effective field theories which respect the symmetries
of the low energy physics.  
In this paper, we follow the later approach and assume that the Standard Model is a good approximation
to physics at the weak scale and that all new physics is at a high scale. 
If the new physics is at a scale much higher than that 
probed experimentally, then an effective theory can be written in terms of an expansion in
higher dimension operators,
\beq
\call\sim {\call}_{SM}+\Sigma_i
{f_i\over \Lambda^2}{\cal{O}}_i + . . . \, .
\label{leff}
\eeq
The lowest dimension operators, ${\cal O}_i$, which contribute to processes involving gauge bosons
are dimension-6.
The complete set of $SU(3)\times SU(2)_L\times U(1)_Y$ operators is quite
 large\cite{Buchmuller:1985jz,Grzadkowski:2010es},
but the assumption of flavor conservation,
 CP conservation, and Standard Model (SM) physics in the lightest $2$ 
fermion generations reduces
the set of allowed operators considerably{\cite{Han:2004az}.
In this work, we further restrict the set of
operators  to those involving only the Higgs  doublet and electroweak  boson 
sectors\cite{Hagiwara:1993ck,Hagiwara:1993qt,Bagger:1992vu,Alam:1997nk,GonzalezGarcia:1999fq}. 
The Lagrangian of Eq. \ref{leff} is valid at low energy ($\sim m_Z$) and reflects our ignorance of high scale
physics. 

At each order in the
expansion in $1/\Lambda^{n}$, divergences appear in loop diagrams, which can
be absorbed by renormalizing the coefficients of the operators appearing at one lower level
in the loop expansion but  at the same order in $1/\Lambda^{n}$,
yielding a theory that is finite order by order in the expansion in $1/\Lambda$.   The low energy effective
field theory has the advantage that it can be matched to many possibilities for high scale
physics. 
The Lagrangian
of Eq. \ref{leff}  can be used at energy scales much below $\Lambda$, where the
observed physics approximates the SM up to small corrections.   A specific model of new physics
will predict the coefficients at the high scale, $f_i(\Lambda)$.  The predictions must then be matched
with experimental limits from low energy measurements, $f_i(\sim m_Z)$, and the renormalization group
used to run the coefficients from the scale of the high energy predictions to  that of the lower energy 
measurements\cite{Grojean:2013kd,Jenkins:2013wua,Jenkins:2013zja,Elias-Miro:2013mua,Elias-Miro:2013gya}. 

New  physics in the electroweak sector can be restricted by measurements of the 
oblique parameters, $\Delta S, ~\Delta T$, and $\Delta U$, 
by restrictions on the deviations of the three gauge boson vertices  and Higgs branching
ratios from the Standard Model predictions, along with many SM processes.
These effects and
the resulting limits on the coefficients, $f_i$,  of the 
dimension $-6$ operators have been studied
by many 
authors\cite{GonzalezGarcia:1999fq,Corbett:2012dm,Masso:2012eq,Contino:2013kra,Dumont:2013wma,Alloul:2013naa,Pich2013:fea,Baak:2013fwa,Banerjee:2013apa,Chang:2013cia}.
 The new feature of our work is a
careful study of the dependence of the predictions on the renormalization scheme
required by the effective theory formalism\cite{Hagiwara:1993ck,Mebane:2013zga,Mebane:2013cra}.
A complete analysis would include fermion operators and a careful choice of basis.  Such a study
is not necessary to illustrate our main point, however, which concerns the numerical effects
of the renormalization scheme in the effective framework. 

In Sec. \ref{sec_eff}, we review the low energy effective Higgs electroweak theory and make the
connection between our notation and the commonly used strongly interacting light Higgs (SILH) 
basis\cite{Giudice:2007fh}.
Section \ref{sec:calc} contains analytic results for the
oblique parameters, along with a discussion of the pinch technique needed to obtain gauge invariant 
results. 
Detailed appendices contain analytic results for the contributions to $\Delta S,~\Delta T$ and  $\Delta U$,
as well as the required pinch contributions, 
in an $R_\xi$ gauge.
 Numerical fits to the coefficients of the effective operators resulting  from the oblique parameters 
and a discussion of the dependence on the renormalization scheme
used to render the effective field theory finite are given in Sec. \ref{sec:phen}.  
As pointed out in Refs. \cite{Mebane:2013zga,Mebane:2013cra},
 our limits are considerably 
weaker than those obtained using only the contributions to the oblique
parameters proportional to $\log(\Lambda)$ given 
in Ref. \cite{Alam:1997nk}.
We also compare the restrictions on the $f_i$
couplings from the oblique parameters with those obtained
 from the the deviations of the experimental results from Standard Model predictions
of Higgs branching ratios.  Finally, Sec. \ref{sec:conc}
contains some conclusions.

\section{Effective Theory}
\label{sec_eff}
The effective Lagrangian we consider contains the $SU(2)_L\times U(1)_Y$ gauge fields and the Higgs
$SU(2)_L$ doublet $\Phi$.  We assume that fermion interactions with the gauge bosons are those given
by the Standard Model and  that all new physics respects the $SU(2)_L\times U(1)_Y$ gauge 
invariance and conserves $C$ and $P$.  Possible new physics effects in the fermion sector are not
considered here. 

There are $4$ operators which affect the gauge boson $2-$point functions at tree 
level\cite{Longhitano:1980tm,Hagiwara:1993ck},
\begin{eqnarray}
{\cal{O}}_{DW}&= & -{g^2\over 4} Tr\biggl(\biggl[D_\mu, \sigma^a \cdot W^a_{\nu \rho}\biggr]
\biggl[D^\mu, \sigma^b\cdot W^{b, \nu\rho}\biggr]\biggr)\nonumber \\
{\cal{O}}_{DB}&= & -{g^{\prime 2}\over 2}(\partial_\mu B_{\nu\rho})
(\partial^\mu B^{\nu\rho})\nonumber \\
{\cal{O}}_{BW}&= &-{g g^\prime \over 4}\Phi^\dagger B_{\mu\nu}\sigma^a 
\cdot W^{a,\mu\nu}\Phi \nonumber \\
{\cal{O}}_{\Phi, 1}&= & (D_\mu\Phi)^\dagger(\Phi \Phi^\dagger)(D^\mu \Phi)\, ,
\label{ops_0l}
\end{eqnarray}
where,\footnote{The convention for $D_\mu$ differs from that of Refs. \cite{Alam:1997nk,Hagiwara:1993ck,GonzalezGarcia:1999fq}
 leading to some minus signs in the literature,
relative to ours, which are described in Appendix A.}
\begin{eqnarray}
D_\mu&=&\partial_\mu-i{g\over 2}B_\mu-i{g^\prime \sigma^2\over 2}W_\mu^a
\nonumber \\
B^{\mu\nu}&=  &
\partial_\mu B_\nu-\partial_\nu B_\mu
\nonumber \\
W^{\pm,\mu\nu}&= & \partial_\mu W_\nu^\pm-\partial_\nu W_\mu^\pm \mp
 i g (W_\mu^3 W_\nu^\pm -W_\nu^3 W_\mu^\pm)
\nonumber \\
W^{3,\mu\nu}&= & \partial_\mu W_\nu^3-\partial_\nu W_\mu^3 -
 i g (W_\mu^+ W_\nu^- -W_\nu^+ W_\mu^-)\, .
\end{eqnarray}
${\cal{O}}_{BW}$ gives $B-W^3$ mixing at tree level and contributes to $\Delta S$, while
${\cal {O}}_{\Phi 1}$ affects $m_Z$, but not  $m_W$, at tree level and thus contributes
to $\Delta T$.  

There are $6$ bosonic operators which contribute to the oblique parameters at $1-$loop,
\begin{eqnarray}
{\cal{O}}_{WWW}&=&-i{g^3\over 8}Tr\biggl(\sigma^a\cdot W^{a,\mu}_\nu \sigma^b\cdot
W^{b,\nu}_\rho\sigma^c\cdot W^{c,\rho}_\mu\biggr)\nonumber \\
{\cal{O}}_{W}&=&i{g\over 2}
   (D_\mu \Phi)^\dagger \sigma^a\cdot W^{a,\mu\nu} (D_\nu \Phi)\nonumber \\
{\cal{O}}_{B}&=&i{g^\prime\over 2}
   (D_\mu \Phi)^\dagger B^{\mu\nu} (D_\nu \Phi)\nonumber\\
{\cal{O}}_{WW}&=&-{g^2\over 4}\Phi^\dagger \sigma^a\cdot W^{a,\mu\nu}\sigma^b
\cdot W^b_{\mu\nu} \Phi\nonumber \\
{\cal{O}}_{BB}&=&-{g^{\prime 2}\over 4}\Phi^\dagger B^{\mu\nu}B_{\mu\nu}\Phi\,\nonumber \\
{\cal{O}}_{\Phi, 2}&= & {1\over 2} \partial^\mu (\Phi^\dagger \Phi)\partial_\mu (
\Phi^\dagger\Phi).
\label{ops_1l}
\end{eqnarray}
The effects of ${\cal O}_{BB}$ and ${\cal O}_{WW}$ on 3-gauge boson vertices can be eliminated
by wave function renormalization and coupling redefinition, leaving only contributions to effective $VVH$
vertices.  

In addition, there are $2$ operators that are neglected in our analysis,
\begin{eqnarray}
{\cal{O}}_{\Phi, 3}&=& {1\over 3}(\Phi^\dagger\Phi)^3 \nonumber \\
{\cal{O}}_{\Phi, 4}&=& (D_\mu\Phi)^\dagger(D^\mu \Phi)(\Phi^\dagger\Phi)\,.
\end{eqnarray}
${\cal{O}}_{\Phi,3}$ only affects the Higgs self interactions, and is irrelevant for the oblique parameters.  
By using the equations of motion, ${\cal{O}}_{\Phi,4}$ can be written as a linear combination of ${\cal{O}}_{\Phi,2}$ 
and dimension-$6$ Yukawa operators.  The latter have no effect on the self energies of gauge bosons, and thus 
${\cal{O}}_{\Phi,4}$ can also be excluded from our operator basis.

Finally, the Lagrangian for Higgs physics we consider is,
\begin{eqnarray}
{\cal L}&=&{\cal L}_{SM}
+{f_{DW}\over \Lambda^2}{\cal O}_{DW}
+ {f_{DB}\over \Lambda^2}{\cal O}_{DB}
+{f_{BW}\over \Lambda^2}{\cal O}_{BW}
+{f_{\Phi, 1}\over \Lambda^2}{\cal O}_{\Phi, 1}
+{f_{\Phi, 2}\over \Lambda^2}{\cal O}_{\Phi, 2}
\nonumber \\
&&
+{f_{WWW}\over \Lambda^2}{\cal O}_{WWW}
+{f_{W}\over \Lambda^2}{\cal O}_{W}
+{f_{B}\over \Lambda^2}{\cal O}_{B}
+{f_{WW}\over \Lambda^2}{\cal O}_{WW}
+{f_{BB}\over \Lambda^2}{\cal O}_{BB}\, .
\label{lag_eff}
\end{eqnarray}
The relationship between the coefficients of Eq. \ref{lag_eff} and those of the SILH Lagrangian are
given in Appendix A for convenience.

\section{Results}
\label{sec:calc}
We are interested in the contributions to the oblique parameters from the operators of
Eq. \ref{lag_eff}.  The computation requires both the gauge boson $2-$ point functions 
and the pinch contributions in order to get gauge invariant results\cite{Degrassi:1993kn}. 
The $2-$ point functions are defined as,
\begin{equation}
\Pi_{XY}^{\mu\nu}(q^2)=g^{\mu\nu}\Pi_{XY}(q^2)-p^\mu p^\nu B_{XY}(q^2)\, ,
\end{equation}
for $XY=WW,ZZ,\gamma\gamma$ and $Z\gamma$.  The contributions 
(including tadpole diagrams) 
to $\Pi_{XY}(q^2)$ are given in Appendix B in $R_\xi$ gauge.\footnote{Results in the unitary gauge can be found in  Ref. \cite{Dawson:1994fa}. } The $\Pi_{XY}$ functions are gauge dependent and ultraviolet divergent. 

The pinch contributions are defined in terms of the corrections to the ${\overline f}
\gamma^\mu P_L f^\prime V^\mu$ vertices, where $P_L={1\over 2}(1-\gamma_5)$.
Only left-handed contributions arise because there is always a coupling to a $W$
boson.
The vertex corrections are of the form,
\begin{equation}
\Delta\Gamma_\mu^{Vff^{\prime}}(q^2)=
\gamma _\mu P_L\Delta \Gamma_L^{Vff^{\prime}}(q^2)
\end{equation}
We normalize,
\begin{eqnarray}
\Delta \Gamma_L^{Vff}(q^2)&=&gT_3^f\Delta \Gamma_L^V(q^2)\quad {\hbox{V=Z}},\gamma
\nonumber \\
\Delta \Gamma_L^{Wff^\prime}(p^2)&=&{g\over \sqrt{2}}
\Delta \Gamma_L^W(q^2)\quad {\hbox{V=W}}\, .
\label{pinchdef}
\end{eqnarray}
Expressions for the pinch contributions,
$\Delta \Gamma_L^{{V}}(q^2)$,
in $R_\xi$ gauge are given in Appendix C.

Gauge invariant $2-$ point functions can be constructed by forming the 
combinations,\footnote{Gauge invariant expressions for the $2-$point functions are found
in Refs. \cite{Mebane:2013zga,Mebane:2013cra}.  Their construction differs from ours by  finite, gauge invariant
 terms, as explained in
Appendix C.}
\begin{eqnarray}
{\overline{\Pi}}_{WW}(q^2)&=&
\Pi_{WW}(q^2)+2(q^2-m_W^2)\Delta\Gamma_L^W(q^2)\nonumber \\
{\overline{\Pi}}_{ZZ}(q^2)&=&
\Pi_{ZZ}(q^2)+2c (q^2-m_Z^2)\Delta\Gamma_L^Z(q^2)\nonumber \\
{\overline{\Pi}}_{\gamma Z}(q^2)&=&
\Pi_{\gamma Z}(q^2)
+s q^2 \Delta\Gamma_L^Z(q^2)+c(q^2-m_Z^2)
\Delta\Gamma_L^\gamma(q^2) \nonumber \\
{\overline{\Pi}}_{\gamma\gamma}(p^2)&=&
\Pi_{\gamma \gamma}(p^2)+2 s q^2 \Delta\Gamma_L^\gamma (q^2)\, ,
\end{eqnarray}
where $c\equiv \cos\theta_W$ and $s\equiv \sin\theta_W$.
We have explicitly verified the cancellation of the $\xi$ gauge parameters.

This allows the construction of gauge invariant oblique parameters\cite{Degrassi:1993kn},
\begin{eqnarray}
\alpha \Delta S&=&
\biggl({4 s^2 c^2\over m_Z^2}\biggr)
\biggl\{ {\overline{\Pi}}_{ZZ}(m_Z^2)- 
{\overline{\Pi}}_{ZZ}(0)-
{\overline{\Pi}}_{\gamma\gamma}(m_Z^2)
\nonumber \\ &&
-{c^2-s^2\over c s}\biggl(
{\overline{\Pi}}_{\gamma Z}(m_Z^2)
\biggr)\biggr\}\nonumber \\
\alpha  \Delta T &=&
 \biggl({ 
 {\overline{\Pi}}_{WW}(0)\over m_W^2}-{
{\overline{\Pi}}_{ZZ}(0)\over m_Z^2}
\biggr)
\nonumber \\
\alpha \Delta U&=& 4 s ^2\biggl\{
{ 
{\overline{\Pi}}_{WW}(m_W^2)-
{\overline{\Pi}}_{WW}(0)\over m_W^2} -c^2
\biggl({ {\overline{\Pi}}_{ZZ}(m_Z^2)-
{\overline{\Pi}}_{ZZ}(0)\over m_Z^2}\biggr)
\nonumber \\
&&-2 s c
\biggl(
{ 
{\overline{\Pi}}_{\gamma Z}(m_Z^2)
\over m_Z^2}\biggr)
-s ^2
{ {\overline{\Pi}}_{\gamma \gamma}(m_Z^2)\over m_Z^2}\biggr\} \, .
\label{sdef}
\end{eqnarray}
The oblique parameters effectively capture the new physics consequences on low energy observables with the
assumption that all new physics is at a scale much larger than $M_Z$ and that the new physics contributes only   to
the 2-point functions.  The non-pinch terms which contribute to the vertex function are hence not included here, but 
generate additional contributions from the effective Lagrangian which is not captured in the STU formalism.
We parameterize the oblique parameters  as,
\begin{eqnarray}
\Delta S &=& C_S{1\over\epsilon}
\biggl({
4\pi\mu^2\over m_Z^2}\biggr)^\epsilon\Gamma(1+\epsilon)+R_S\nonumber \\
\Delta T &=& C_T{1\over\epsilon}
\biggl({
4\pi\mu^2\over m_Z^2}\biggr)^\epsilon
\Gamma(1+\epsilon)
+R_T\nonumber \\
\Delta U &=& C_U{1\over\epsilon}
\biggl({
4\pi\mu^2\over m_Z^2}\biggr)^\epsilon
\Gamma(1+\epsilon) +R_U\, ,
\label{genans}
\end{eqnarray}
where
\begin{eqnarray}
C_S&=&{m_H^2\over 8 \pi}\biggl\{ {f_B+f_W\over \Lambda^2}\biggr\}+{m_Z^2\over 24\pi\Lambda^2}
\biggl\{ f_B(20c^2+7)- 3 f_W
\nonumber \\ &&
+24(s^2f_{BB}+c^2f_{WW})+36c^2g^2f_{WWW} + {8 c^2 \over g^2} f_{\Phi,2}\biggr\}\nonumber \\
C_T &=&{1\over 16\pi c^2}\biggl\{ 9m_W^2\biggl({f_B+f_W\over \Lambda^2}\biggr)
+3m_H^2 {f_B\over \Lambda^2} -12{m_W^2\over g^2}{f_{\Phi,2}\over \Lambda^2} \biggr\}
 \nonumber \\
C_U &=&{m_Z^2\over 6\pi \Lambda^2}s^2f_W \,.
\label{cdef}
\end{eqnarray}
The logarithmic contributions to the oblique parameters coefficients have been obtained   in 
Ref. \cite{Alam:1997nk} and are
obtained by the replacement in Eq. \ref{genans},
\begin{equation}
{1\over\epsilon}
\biggl(
{4\pi\mu^2\over m_Z^2}
\biggr)^\epsilon\Gamma(1+\epsilon) \rightarrow \ln\biggl({\Lambda^2\over m_Z^2}\biggr)\, .
\end{equation}
All other terms are dropped, which gives the leading logarithmic result,
\begin{eqnarray}
\Delta S^{LL}&\rightarrow&C_S\log\biggl({\Lambda^2\over m_Z^2}\biggr)
\nonumber \\
\Delta T^{LL}&\rightarrow&C_T\log\biggl({\Lambda^2\over m_Z^2}\biggr)
\nonumber \\
\Delta U^{LL}&\rightarrow&C_U\log\biggl({\Lambda^2\over m_Z^2}\biggr)\, .
\label{obl_ll}
\end{eqnarray}

We take as inputs,
\begin{eqnarray}
&m_H=126~GeV, \quad &m_Z=91.1875~GeV,\quad m_W=80.399~GeV\nonumber\\
&G_F=1.16637\times 10^{-5}~GeV^{-2}, \quad &m_t=173~GeV\, .
\end{eqnarray}
All other inputs are obtained using the tree level relationships of the SM. Eq. \ref{obl_ll} becomes,
\begin{eqnarray}
\Delta S^{LL}&=&  \biggl[
0.015 f_B
+.0014 f_W
+.0028 f_{BB}
+.01f_{WW}+.006f_{WWW} +.0016 f_{\Phi,2}
\biggr]
\nonumber \\
&&\cdot
\biggl({1~TeV\over \Lambda}\biggr)^2
\biggl[
{\log(\Lambda^2/m_Z^2)
\over  \log(1~TeV^2/ m_Z^2)}\biggr]
 \nonumber\\
\Delta T^{LL}&=&\biggl[
0.013 f_B+.007 f_W - .0047 f_{\Phi,2}
\biggr]
\biggl({1~TeV\over \Lambda}\biggr)^2   
\biggl[
{\log(\Lambda^2/m_Z^2)
\over  \log(1~TeV^2/ m_Z^2)}\biggr]
 \nonumber \\
\Delta U^{LL}&=&
\biggl[
0.0005f_W
\biggr]
\biggl({1~TeV\over \Lambda}\biggr)^2  
\biggl[
{\log(\Lambda^2/m_Z^2)
\over  \log(1~TeV^2/ m_Z^2)}\biggr]
 \nonumber \\
\label{numll}
\end{eqnarray}
Numerical fits to the oblique parameters using  the logarithmic
contributions of Eq. \ref{numll}  have been given in Refs. \cite{Corbett:2012dm,Corbett:2012ja}.  
The terms not associated with the divergences are,\footnote{Since $f_{BW}, f_{\Phi,1}$, and $f_{DW}$ contribute at tree level, we do not consider the much smaller contributions of these operators at $1$-loop to the oblique parameters.}
\begin{eqnarray}
R_S&=&-{4\pi v^2\over \Lambda^2} f_{BW} + 
 R_{S1}
+R_{S2}\log(c)+R_{S3}\log\biggl({m_H\over m_Z}\biggr)\nonumber \\
R_T&=&-{v^2\over 2 \alpha \Lambda^2}f_{\Phi,1}+ R_{T1}
+R_{T2}\log(c)+R_{T3}\log\biggl({m_H\over m_Z}\biggr)\nonumber \\
R_U&=&{g^2 s^2\over c^2}{8\pi v^2\over \Lambda^2} f_{DW}
+ R_{U1}
+R_{U2}\log(c)+R_{U3}\log\biggl({m_H\over m_Z}\biggr)\, .
\end{eqnarray}
Analytic results for $R_S,~R_T,$ and $R_U$ are given in Appendix D.
Numerically we find,
\begin{eqnarray}
R_{S}&=& \biggl\{ -0.76 f_{BW}+ 10^{-3} \bigl(1.48 f_B - 1.4 f_W-0.2 f_{BB}-0.71 f_{WW} \nonumber \\
&& + 0.66 f_{WWW}+ 1.96 f_{\Phi,2}\bigl) \biggr\}\biggl({1~TeV\over \Lambda}\biggr)^2 \nonumber \\
R_{T}&=& \biggl\{-4.0 f_{\Phi,1} - 10^{-3} \bigl(0.13 f_B+0.12 f_W - 3.97 f_{\Phi,2}\bigr)
\biggr\}\biggl({1~TeV\over \Lambda}\biggr)^2
\nonumber \\
R_{U}&=& \biggl\{ 0.20 f_{DW} + 10^{-3} \bigl(-0.02 f_B + 2.06 f_W+ 0.14 f_{WW} \nonumber \\
&&+ 2.1 f_{WWW}- 0.25 f_{\Phi,2}\bigr)
\biggr\}\biggl({1~TeV\over \Lambda}\biggr)^2 .
\label{rennumb}
\end{eqnarray}
The effective field theory is defined at the weak scale, $\mu\sim m_Z$, and encapsulates the effects of
potential new physics which may occur at high scales.  
The divergences which arise at $1-$ loop, Eq. \ref{genans}, can be eliminated by renormalizing the
coefficients which enter the tree level results. 
The theory is thus rendered  finite order by order in 
the expansion in powers of $1/\Lambda^2$\cite{Mebane:2013zga,Mebane:2013cra,Bagger:1992vu,Hagiwara:1993ck,Einhorn:2013tja}.
 Using ${\overline {MS}}$ renormalization,
the renormalized coefficients relevant for a study of the oblique parameters are,
\begin{eqnarray}
f_{BW}(\mu)&=&f_{BW}-{1\over \epsilon}(4\pi)^\epsilon \Gamma(1+\epsilon) C_S\nonumber \\
f_{DW}(\mu)&=&f_{DW}-{1\over \epsilon}(4\pi)^\epsilon \Gamma(1+\epsilon) C_U\nonumber \\
f_{\Phi,1}(\mu)&=&f_{\Phi,1}-{1\over \epsilon}(4\pi)^\epsilon \Gamma(1+\epsilon) C_T\, .
\label{renorm}
\end{eqnarray}
This renormalization prescription is equivalent to that of  \cite{Hagiwara:1993ck}.
The large logarithms of Ref. \cite{Alam:1997nk}
have been eliminated by the renormalization of the
tree level couplings and the only remaining
contributions to the oblique corrections are the finite contributions.  Our final result, 
taking $\mu= m_Z$, as appropriate for the low energy effective Lagrangian, is,
\begin{eqnarray}
\Delta S&=& R_S\nonumber \\
\Delta T&=& R_T\nonumber \\
\Delta U&=& R_U\, .
\label{obfin}
\end{eqnarray}
Note that this result is quite different from that of Ref. \cite{Alam:1997nk},  since the $\log(\Lambda)$ terms
have all been cancelled by the renormalization of the tree level coefficients. 
This is in agreement with the leading result of Ref. \cite{Grojean:2013kd}  for $\Delta S$ which was
obtained by scaling the coefficients in a BSM model from $\Lambda$ to
$m_Z$.

\section{Phenomenology}
\label{sec:phen}
\subsection{Results from Fits to Oblique Parameters}

We do a $\chi^2$ fit to the oblique parameters, using the results of the GFITTER group\cite{Baak:2012kk},
\begin{eqnarray}
\Delta S&=& 0.03\pm 0.10\nonumber \\
\Delta T&=& 0.05 \pm 0.12\nonumber \\
\Delta U &=& 0.03 \pm 0.10
\end{eqnarray}
with the correlation matrix,
\begin{equation}
\rho=\left(\begin{matrix}
1.0&0.891 & -0.540\\
0.891& 1.0 & -0.803\\
-0.540&-0.803&1.0
\end{matrix}\right)\, .
\end{equation}

The parameters $f_{BW}$ and $f_{\Phi,1}$ contribute to $\Delta S$ and $\Delta T$ at tree level.
We show the limits in Fig. \ref{fig_phi1_fbw}.\footnote{ Ref. \cite{Ciuchini:2013pca} has done a similar analysis 
and our fit to $f_{BW}$ and $f_{\Phi,1}$ agrees with theirs.}  These coefficients are highly restricted by the electroweak
data and we ignore them in the remaining fits, where we obtain limits pairwise
on various coefficients.  We do not perform a global fit to the coefficients, since 
our point is simply to illustrate the numerical effects of the renormalization.
Taking the $95\%$ confidence level limits, and 
assuming $f_{BW}(m_Z)$ and $f_{\Phi,1}(m_Z)$ are ${\cal O}(1)$, the fit implies $\Lambda > 1.8~TeV$.
Even though they are numerically constrained, $f_{BW}$ and $f_{\Phi,1}$ cannot be
set to $0$ at the beginning, since they  play a critical role in the
renormalization, as seen in Eq. \ref{renorm}\cite{Mebane:2013zga,Mebane:2013cra}. 
 \begin{figure}[tb]
      \includegraphics[width=.45\textwidth,angle=0,clip]{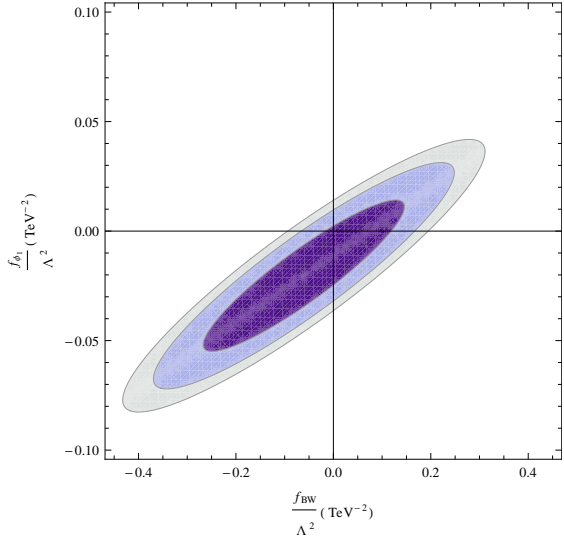}
\caption{Limits from the oblique parameters 
on $f_{\Phi, 1}$ and $f_{BW}$ for $\Lambda=1~TeV$.  These operators contribute
at tree level and are significantly restricted.
 The curves (from outer to inner) are 99, 95, and 68 
$\%$ confidence level.
}
\label{fig_phi1_fbw}
\end{figure}

In Fig. \ref{fig_fbbfww},  we show the allowed region for $\Lambda=1~TeV$ in the 
$f_{BB}(m_Z)$ and $f_{WW}(m_Z)$
plane, setting all other coefficients to zero.  
The left-hand side shows the result using the leading logarithmic
result of Eq. \ref{rennumb}.  This figure is in agreement with Fig. 6 of Ref. \cite{Masso:2012eq}.  After
renormalizing the coefficients of the effective theory, as in Eq. \ref{obfin}, the result is shown on the
right-hand side of Fig. \ref{fig_fbbfww}.  
In Fig. \ref{fig_fwfww} we show the allowed region for $\Lambda=1~TeV$ in the 
$f_{W}(m_Z)$ and $f_{WW}(m_Z)$
plane, setting all other coefficients to zero.  
Again, the left-hand side shows the result using the leading logarithmic
result of Eq. \ref{rennumb}, while the right-hand side shows the
result after renormalization of the couplings.  As emphasized in
Refs. \cite{Mebane:2013zga,Mebane:2013cra}, 
the limits are considerably
weakened once the renormalization procedure, Eq. \ref{renorm}, is applied.   In fact, taking $f_i\sim 1$, we see that no useful limits can be inferred.

 \begin{figure}[tb]
\subfigure[]{
      \includegraphics[width=.45\textwidth,angle=0,clip]{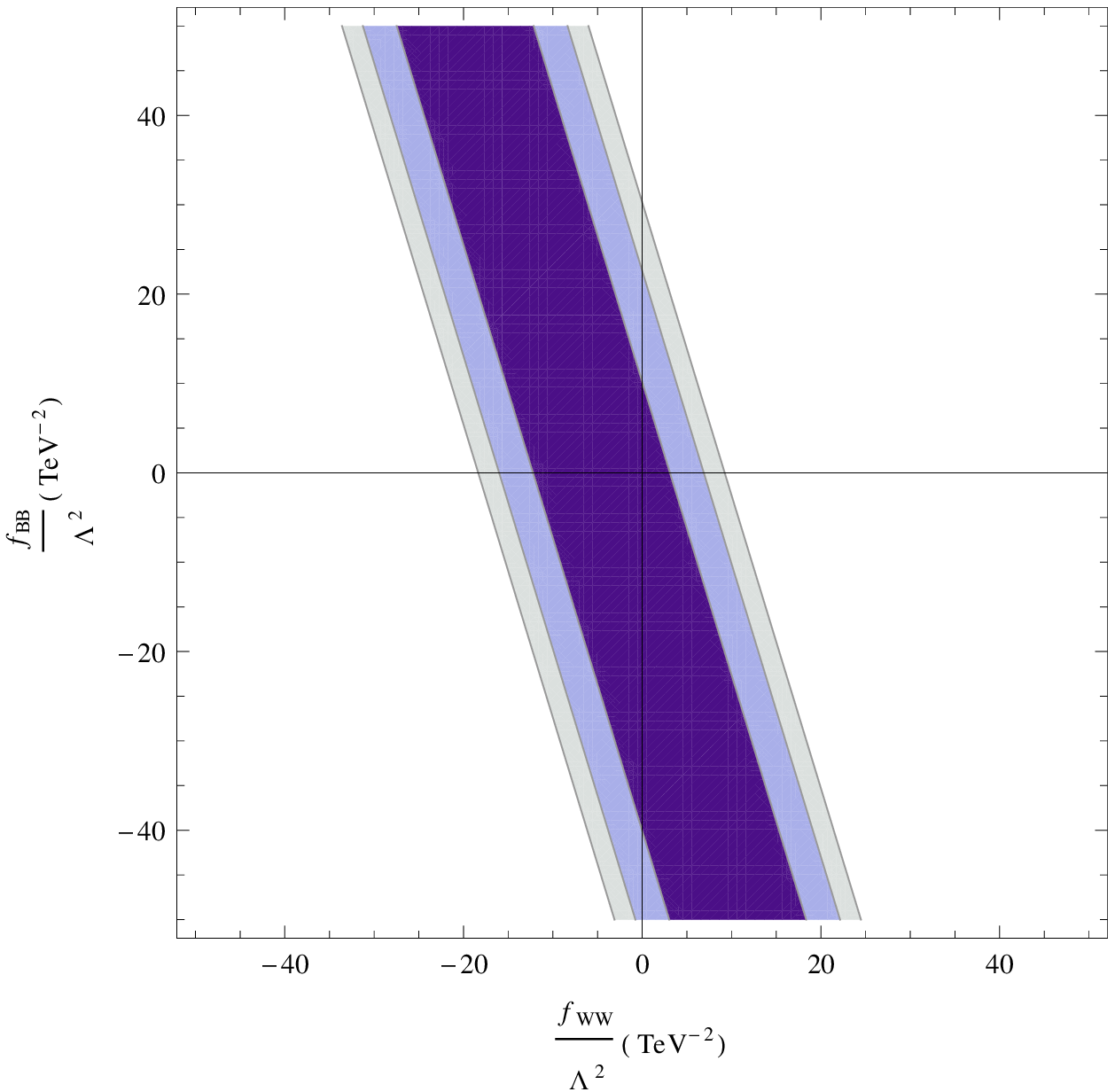}
      }
\subfigure[]{
      \includegraphics[width=0.45\textwidth,clip]{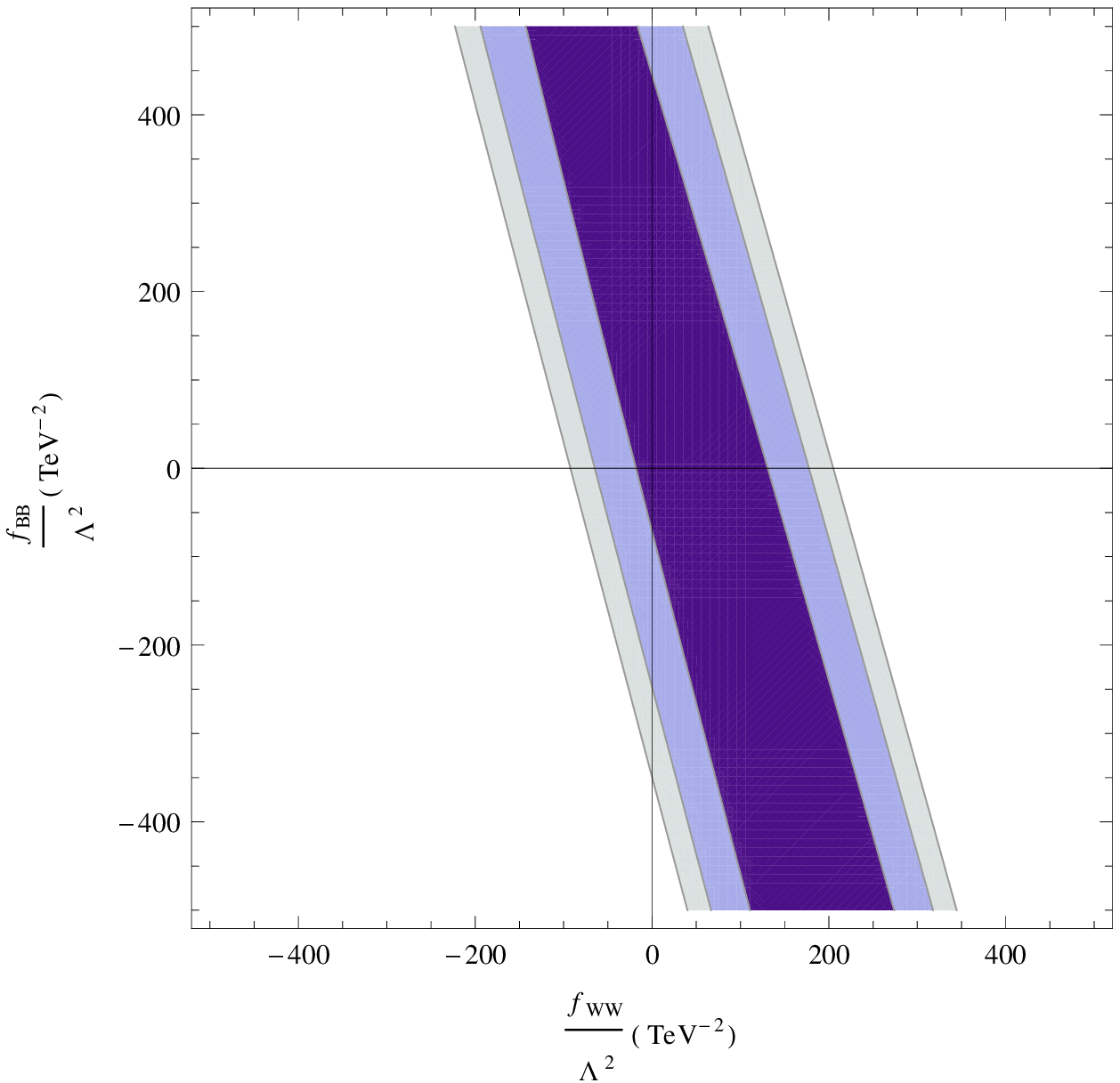}
      }
\caption{(a) Limits from the oblique parameters 
on $f_{BB}$ and $f_{WW}$ for $\Lambda=1~TeV$, using the leading
logarithmic results of Eq. \ref{numll}.  The curves (from outer to inner) are 99, 95, and 68 
$\%$ confidence level. (b) Same as (a) except using the renormalized values of 
the coefficients, $f_{BB}(m_Z)$ and $f_{WW}(m_Z)$, Eqs. \ref{rennumb} and \ref{obfin}.
}
\label{fig_fbbfww}
\end{figure}

 \begin{figure}[tb]
\subfigure[]{
      \includegraphics[width=.45\textwidth,angle=0,clip]{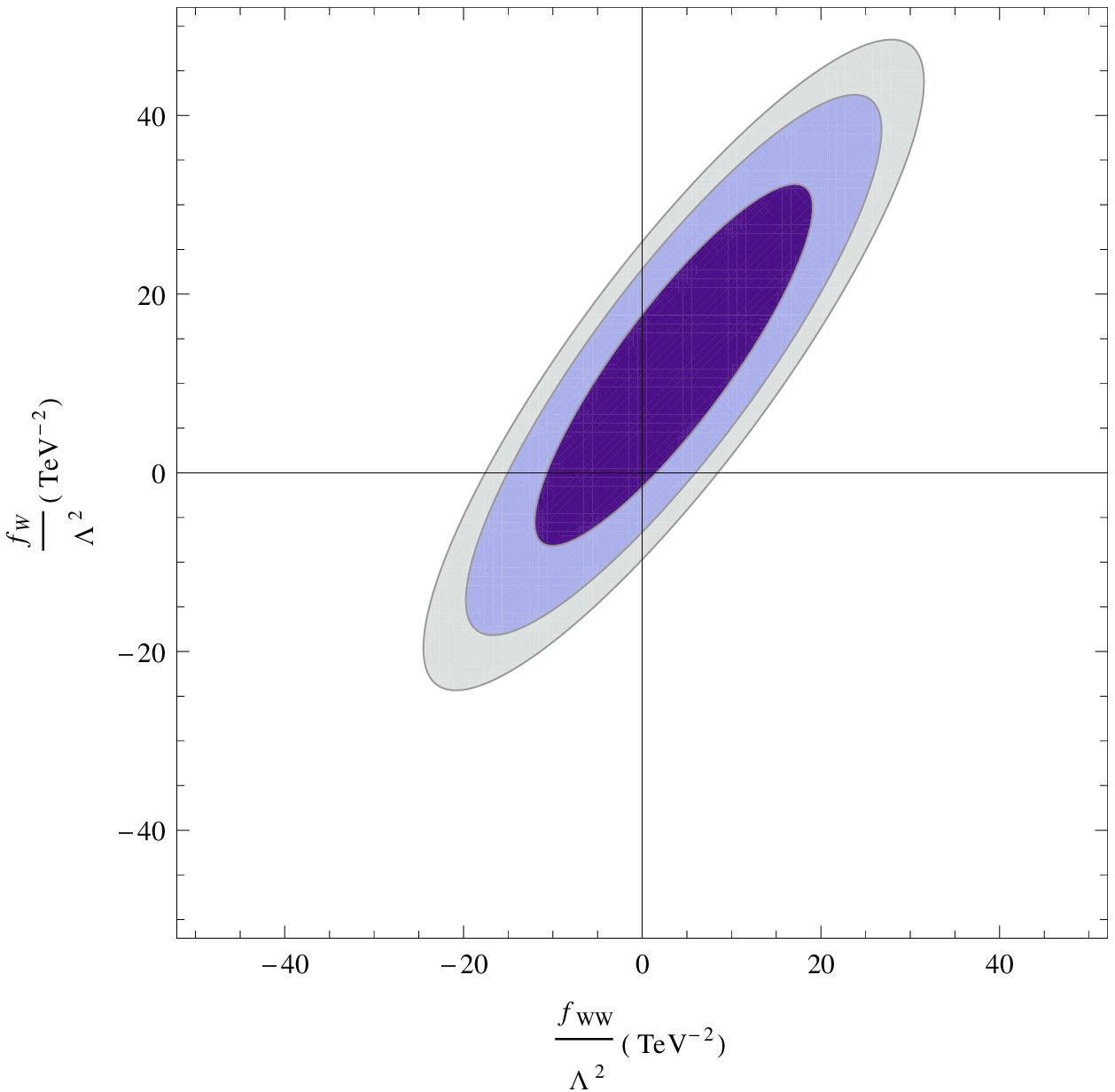}
      }
\subfigure[]{
      \includegraphics[width=0.45\textwidth,clip]{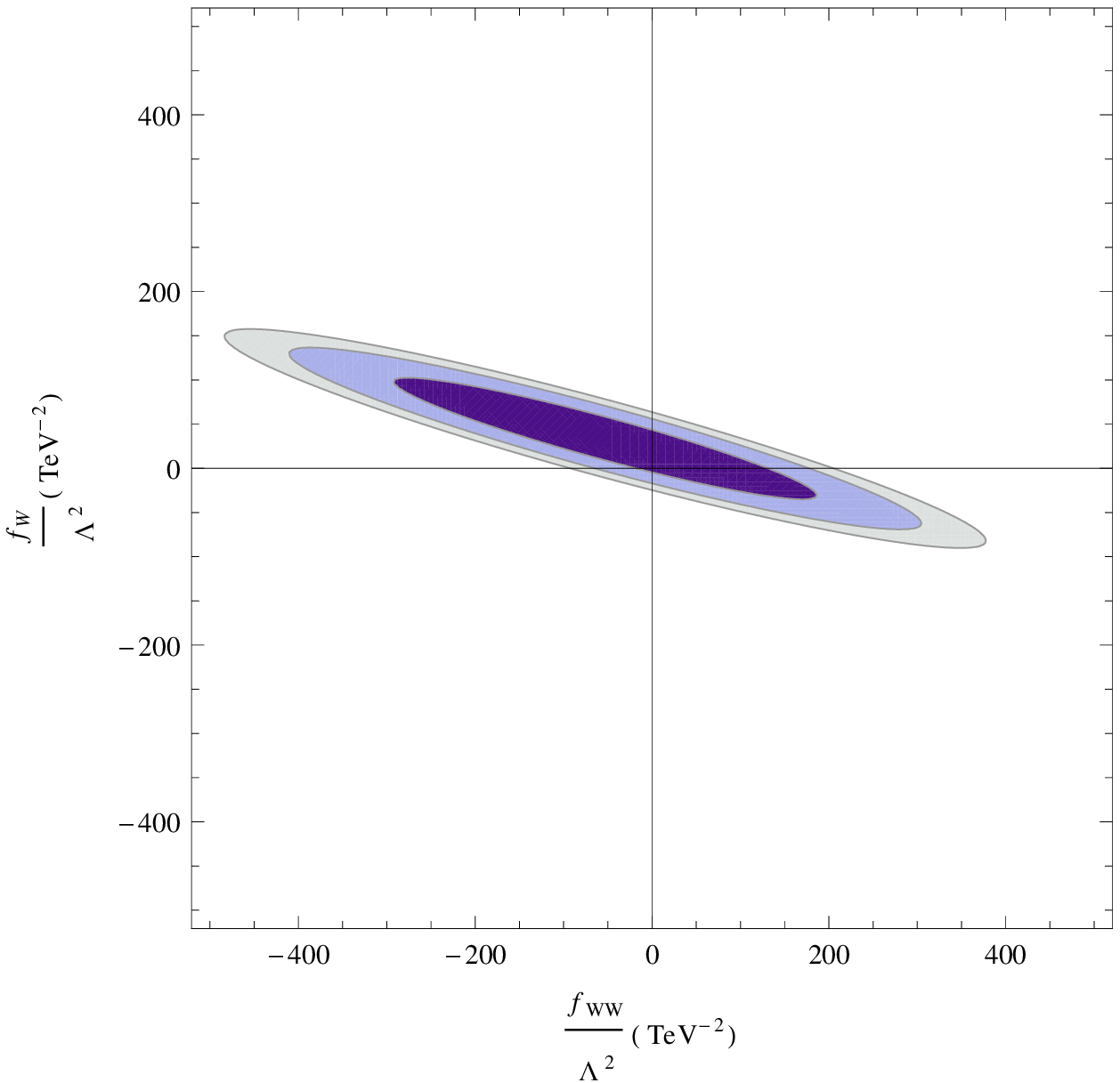}
      }
\caption{(a) Limits from the oblique parameters 
on $f_{W}$ and $f_{WW}$ for $\Lambda=1~TeV$, using the leading
logarithmic results of Eq. \ref{numll}.  The curves (from outer to inner) are 99, 95, and 68 
$\%$ confidence level. (b) Same as (a) except using the renormalized values of 
the coefficients, $f_{W}(m_Z)$ and $f_{WW}(m_Z)$, Eqs. \ref{rennumb} and \ref{obfin}.
}
\label{fig_fwfww}
\end{figure}

\subsection{Implications for Higgs Decays}
In this section, we demonstrate the complementarity of limits from oblique parameters to those
obtained from measurements of Higgs branching ratios. 

In the effective theory, the decay $H\rightarrow W^+W^-$ is 
modified\cite{Hagiwara:1993qt,GonzalezGarcia:1999fq},\footnote{Note our differing convention for the sign of $f_W$ from these references.}
\begin{eqnarray}
\mu_{WW}&\equiv &{\Gamma(H\rightarrow W^+W^-)\over \Gamma(H\rightarrow W^+W^-)\mid_{SM}}
\nonumber \\
&=&{1\over 4 - 4x_W+3x_W^2}
\biggl[
2\biggl(x_W+2f_W(m_Z){m_W^2\over\Lambda^2}
+2f_{WW}(m_Z){m_W^2\over \Lambda^2}(2-x_W)\biggr)^2\nonumber \\
&&
+\biggl(2-x_W+2f_W(m_Z){m_W^2\over\Lambda^2}+2f_{WW}(m_Z){m_W^2\over \Lambda^2}
x_W\biggr)^2\biggr]-{2 \over g^2}{ m_W^2\over \Lambda^2}(f_{\Phi,1}+2 f_{\Phi,2})\nonumber \\ 
&\sim & 1+\biggl[.0086 f_{WW}(m_Z) + .017 f_{W}(m_Z) \nonumber \\
&& \hspace{0.3cm} -.03 f_{\Phi,1}(m_Z)-.06 f_{\Phi,2}(m_Z)
\biggr]\biggl({1~TeV\over \Lambda}\biggr)^2+{\cal O}\biggl({1\over \Lambda^4}\biggr)\,, 
\label{hww}
\end{eqnarray}
where $x_W=4m_W^2/m_H^2$ and
we have made explicit the dependence of the coefficients on the scale.  
The $f_{\Phi,1}$ and $f_{\Phi,2}$ contributions come from the Higgs wave function renormalization.
Since we have assumed that
there are no non-SM corrections to the fermion-Higgs couplings, we can use the
measurements of $H\rightarrow W^+W^-$ from gluon fusion to limit $f_{WW}$ and $f_W$
in Eq. \ref{hww}\cite{Aad:2013wqa,Chatrchyan}:
\begin{eqnarray}
\mu_{WW}&=.68\pm .20 \quad & (CMS)
\nonumber \\
\mu_{WW}&=.99\pm .30\quad &(ATLAS)
\, .
\end{eqnarray}
\begin{figure}[tb]
\subfigure[]{
      \includegraphics[width=.45\textwidth,angle=0,clip]{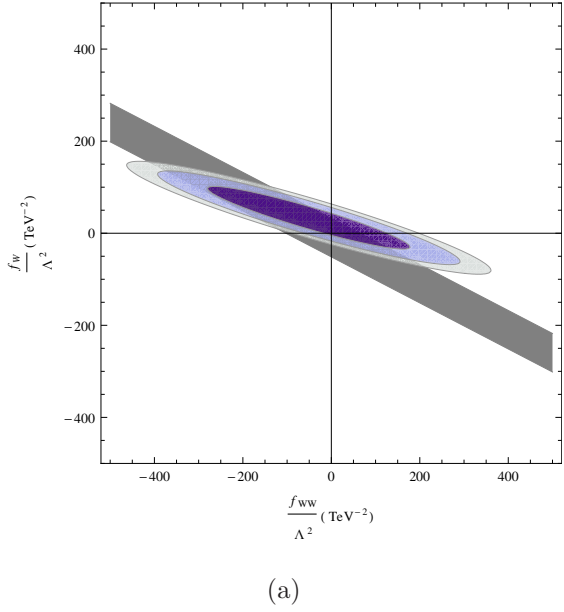}
      }
\caption{ $95\%$ confidence level limits from the measurement of  gluon
fusion production with $H\rightarrow
W^+W^-$ (black band), compared with the inferred limits from the oblique parameters.
}
\label{fig_hww}
\end{figure}
In Fig. \ref{fig_hww}, 
we show the allowed region in $f_{WW}(m_Z)$ versus $f_W(m_Z)$ from $H\rightarrow W^+W^-$. In this case, the limits from Higgs decay and from the oblique parameters are similar.
 Note that the scale of Fig. \ref{fig_hww}, $f_i/\Lambda^2\sim 200$, makes these limits meaningless.

Only $f_{BB}$, $f_{WW}$, and $f_{BW}(m_Z)$ contribute to $H\rightarrow \gamma \gamma$.
Using the well known SM results\cite{Gunion:1989we} we find,
\begin{eqnarray}
\mu_{\gamma\gamma}&\equiv&
{\Gamma(H\rightarrow \gamma\gamma)\over \Gamma(H\rightarrow\gamma
\gamma)\mid_{SM}}
\nonumber \\
&=&
\biggl\{1+
\biggl(
{I_{real}\over I_{real}^2+I_{imag}^2}
\biggr)
{8\pi^2v^2 \over \Lambda^2}
\biggl[
f_{BB}(m_Z)+f_{WW}(m_Z)-f_{BW}(m_Z)\biggr]\biggr\}^2
\nonumber \\
&\sim&
1+1.47\biggl(
{1~TeV \over \Lambda}\biggr)^2
\biggl[
f_{BB}(m_Z)+f_{WW}(m_Z)-f_{BW}(m_Z)\biggr]
+{\cal O}\biggl({1\over\Lambda^4}\biggr)\,
\end{eqnarray}
where
\begin{eqnarray}
I_{real}&=&\Sigma_f N_C Q_f^2F_{1/2}^{real}
(x_f)
+F_{1}^{real}(x_W)
\nonumber \\
I_{imag}&=& \Sigma_f N_C Q_f^2F_{1/2}^{imag}
(x_f)
+F_{1}^{imag}(x_W)\, ,
\end{eqnarray}
$x_f=4m_f^2/m_H^2$
and expressions for $F_{1/2}$ and $F_1$ are found in Ref. \cite{Gunion:1989we}.
We can do a simple fit to the ATLAS and CMS results for $H\rightarrow \gamma\gamma$,
\cite{Aad:2013wqa,Chatrchyan}
\begin{eqnarray}
\mu_{\gamma\gamma}&=.77\pm .27 \quad & (CMS)
\nonumber \\
\mu_{\gamma\gamma}&=1.55\pm .31\quad &(ATLAS)
\, .
\end{eqnarray}
Fig. \ref{fig_hgg} shows the $95\%$ confidence level limits from the 
gluon fusion of the Higgs boson, with the subsequent decay
to $\gamma\gamma$, and contrasts the limit from the oblique parameters,
(setting $f_{BW}(m_Z)=0$).  The error band on the $H\rightarrow \gamma
\gamma$ limits
is not apparent on this scale, and again it is clear that the constraints from the oblique parameters cannot
compete with those from Higgs decay even though they have a different shape in the $f_{WW}$ and $f_{BB}$ planes.
\begin{figure}[tb]
\subfigure[]{
      \includegraphics[width=.45\textwidth,angle=0,clip]{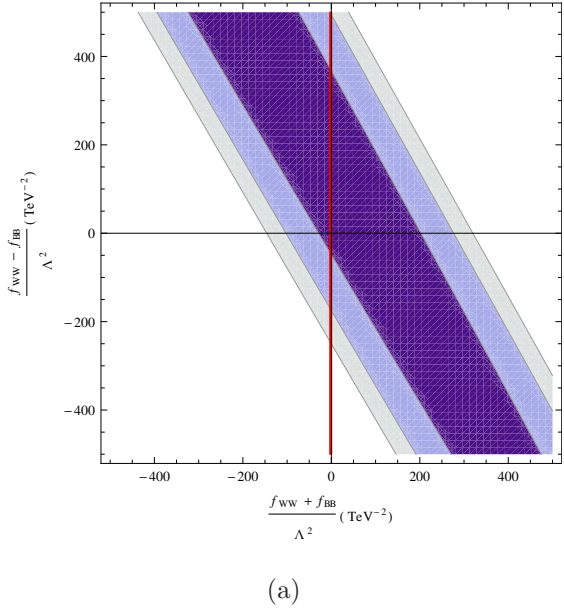}
      }
\caption{ Comparison of limits from oblique parameters (blue bands)
and $95\%$ confidence level limits from the measurement of  gluon
fusion production and the subsequent $H\rightarrow \gamma \gamma$ decay (red vertical line).
}
\label{fig_hgg}
\end{figure}

In a similar fashion, we can find the contribution to $H\rightarrow Z\gamma$,
\begin{eqnarray}
\mu_{Z\gamma}& \equiv &
{\Gamma(H\rightarrow Z \gamma)\over \Gamma(H\rightarrow Z
\gamma)\mid_{SM}} \nonumber \\ 
&=&1+{2A_{real}\over A_{real}^2+A_{imag}^2}{2\pi s c\over \alpha}{m_Z^2\over\Lambda^2}g_1
+{\cal O}\biggl({1\over \Lambda^4}\biggr)
\end{eqnarray}
where\cite{Hagiwara:1993qt}
\begin{equation}
g_1={f_B(m_Z)-f_W(m_Z)+4s^2f_{BB}(m_Z)-4c^2f_{WW}(m_Z)+2(c^2-s^2)f_{BW}(m_Z)}\, ,
\end{equation}
and $A_{real}$ and $A_{imag}$ are the real and imaginary contributions of the sum of the fermion and $W$ loops
to the SM  $H\rightarrow Z\gamma$ decay and can be found in Ref.  \cite{Gunion:1989we}.
In Fig. \ref{fig_hzg}, we show the limits on combinations of $f_i(m_Z)$ which influence the decay $H\rightarrow 
Z\gamma$.  Neglecting all parameters except $f_W(m_Z)$ and $f_B(m_Z)$, Fig. \ref{fig_hzg}
corresponds to 
\begin{equation}
-80<\biggl[f_B(m_Z)-f_W(m_Z)\biggr]\biggl({ 1~TeV\over\Lambda}\biggr)^2 < 35\, ,
\end{equation}
 which is not strong enough to give a limit on $\mu_{Z\gamma}$.

\begin{figure}[tb]
\subfigure[]{
      \includegraphics[width=.45\textwidth,angle=0,clip]{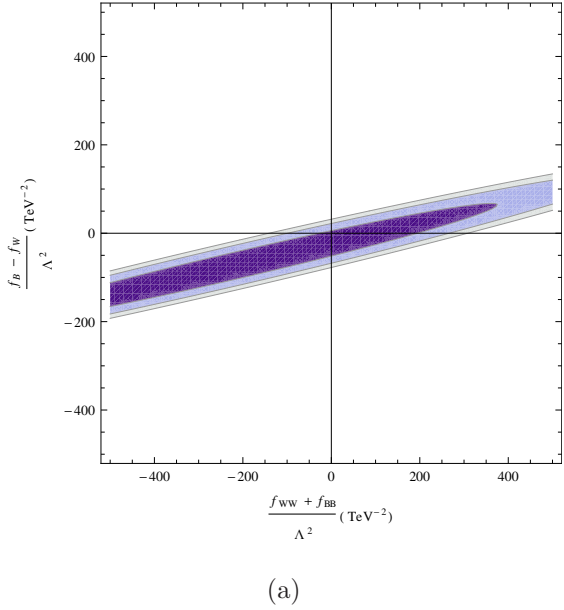}
      }
\caption{Limits from oblique parameters
which influence the decay $H\rightarrow Z \gamma $.  The curves  (from outer to inner) are 99, 95, and 68 
$\%$ confidence level.  
}
\label{fig_hzg}
\end{figure}

\section{Conclusion}
\label{sec:conc}
We have re-examined the prescription of Ref. \cite{Alam:1997nk} 
for obtaining limits on the couplings of an effective low energy 
theory of Higgs-gauge boson interactions by approximating $\Delta S$, $\Delta T$ and $\Delta U$ by the
leading logarithmic contributions.  After renormalizing the coefficients of
the operators which affect tree level results, however, the remaining contributions to the oblique parameters
have no logarithmic enhancement 
and the leading logarithmic approximation is inaccurate.
We give both analytic and numerical results for the oblique parameters.  Fits to the couplings of the effective theory using our prescription for $\Delta S$, $\Delta T$ and $\Delta U$ show that no meaningful limit on the couplings which
contribute at one-loop can be obtained from the oblique parameters. This is in contrast to the couplings which
contribute at tree level which are tightly constrained. 
 This is in agreement with the results of Ref. \cite{Mebane:2013zga,Mebane:2013cra}.  Limits 
on the couplings,
$f_i$ can, however, be extracted from Higgs decays and measurements of 3-gauge boson vertices and a complete fit was given in Ref.  \cite{Corbett:2012ja}. 
\section*{Acknowledgement}
The work of C.-Y. Chen and S. Dawson was supported by the U.S. Department of Energy under DE-AC02-98CH10886.
C. Zhang was supported by  IISN Fundamental Interactions convention 4.4517.08. We thank S. Willenbrock for helpful comments on the manuscript
and M.C. Gonzalez-Garcia for pointing out the need to include ${\cal{O}}_{\Phi, 2}$ in our analysis.
C. Zhang thanks H. Mebane, N. Greiner, and S. Willenbrock for their previous collaboration. 
\newpage
\section*{Appendix A: Relations between HISZ and SILH Basis}
\label{hisz_silh}
%

\newcommand{\gmu}{\gamma^\mu}
\newcommand{\ie}{{\it i.e.}}
\newcommand{\eg}{{\it e.g.}}
\newcommand{\CO}{\mathcal{O}}
\renewcommand{\d}[1]{\ensuremath{\operatorname{d}\!{#1}}}
\newcommand{\myo}[1]{\ensuremath{\mathcal{O}_{#1}}} 
\newcommand{\silh}[1]{\ensuremath{\mathcal{O}^{SILH}_{#1}}}
\newcommand{\myc}[1]{\ensuremath{f_{#1}}} 
\newcommand{\silhc}[1]{\ensuremath{c^{SILH}_{#1}}}


In this Appendix we present a prescription for converting between our effective Lagrangian and 
others which are frequently used in the literature.
We begin by comparing our results with those of Ref. \cite{Hagiwara:1993ck} which we label as HISZ.
Ref. \cite{Hagiwara:1993ck} uses as their convention,
$D^{HISZ}_\mu\equiv\partial_\mu+ig'\frac{1}{2}B_\mu+ig\frac{\sigma^a}{2}W_\mu^a$, 
which is opposite from ours. 
To convert our final results to those of HISZ, 
the substitutions, 
$f_{WWW}\rightarrow -f_{WWW}^{HISZ}$, 
$f_W\rightarrow - f_W^{HISZ}$
and $f_B\rightarrow -f_B^{HISZ}$ must be made in Eq. \ref{genans}.

The HISZ (Hagiwara-Ishihara-Szalapski-Zeppenfeld) operator basis
\cite{Hagiwara:1993ck} has 11 operators involving Higgs and EW gauge fields:
\[
  \myo{DW}^{HISZ}, \myo{DB}^{HISZ}, \myo{BW}^{HISZ}, \myo{\Phi,1}^{HISZ},
  \myo{\Phi,2}^{HISZ}, \myo{\Phi,3}^{HISZ},
  \myo{WWW}^{HISZ}, \myo{WW}^{HISZ}, \myo{BB}^{HISZ}, \myo{W}^{HISZ}, \myo{B}^{HISZ}\ .
\]
The SILH (Strongly-Interacting Light Higgs) operator basis \cite{Giudice:2007fh}
has the same number of operators (involving Higgs and EW gauge fields):
\[
  \silh{H}, \silh{T}, \silh{6},
  \silh{W}, \silh{B}, \silh{2W}, \silh{2B},
  \silh{BB}, \silh{HW}, \silh{HB}, \silh{3W}\ , ,
  \]
where we  use the definition in \cite{Elias-Miro:2013mua}. The
operators are the same, but the normalization factors are more convenient.

The connection between operators are:
\begin{itemize}
  \item From SILH to our convention:
    \begin{flalign}
      \silh{H}&=\myo{\Phi,2}  \nonumber\\
      \silh{T}&=\myo{\Phi,2}-2\myo{\Phi,1}  \nonumber\\
      \silh{6}&=3\lambda\myo{\Phi,3}  \nonumber\\
      \silh{W}&=2\myo{W}-\myo{WW}-\myo{BW}  \nonumber\\
      \silh{B}&=2\myo{B}-\myo{BB}-\myo{BW}  \nonumber\\
      \silh{2W}&=\frac{1}{2g^2}\myo{DW}-\frac{2}{g^2}\myo{WWW}  \nonumber\\
      \silh{2B}&=\frac{1}{2g'^2}\myo{DB}  \nonumber\\
      \silh{BB}&=-4\myo{BB}  \nonumber\\
      \silh{HW}&=2\myo{W}  \nonumber\\
      \silh{HB}&=2\myo{B}  \nonumber\\
      \silh{3W}&=\frac{2}{3g^2}\myo{WWW}  
    \end{flalign}
  \item From our convention to SILH:
    \begin{flalign}
      \myo{DW}&=2g^2\silh{2W}+6g^2\silh{3W} \nonumber\\
      \myo{DB}&=2g'^2\silh{2B} \nonumber\\
      \myo{BW}&=-\silh{B}+\frac{1}{4}\silh{BB}+\silh{HB} \nonumber\\
      \myo{\Phi,1}&=\frac{1}{2}\silh{H}-\frac{1}{2}\silh{T} \nonumber\\
      \myo{\Phi,2}&=\silh{H} \nonumber\\
      \myo{\Phi,3}&=\frac{1}{3\lambda}\silh{6} \nonumber\\
      \myo{WWW}&=\frac{3g^2}{2}\silh{3W} \nonumber\\
      \myo{WW}&=\silh{B}-\silh{W}-\silh{HB}+\silh{HW}-\frac{1}{4}\silh{BB} \nonumber\\
      \myo{BB}&=-\frac{1}{4}\silh{BB} \nonumber\\
      \myo{W}&=\frac{1}{2}\silh{HW} \nonumber\\
      \myo{B}&=\frac{1}{2}\silh{HB} 
    \end{flalign}
\end{itemize}

The connections between coefficients are:
\begin{itemize}
  \item From SILH to our convention:
    \begin{flalign}
      \silhc{H}&=\frac{1}{2}\myc{{\Phi,1}}+\myc{{\Phi,2}} \nonumber\\
      \silhc{T}&=-\frac{1}{2}\myc{{\Phi,1}} \nonumber\\
      \silhc{6}&=\frac{1}{3\lambda}\myc{{\Phi,3}} \nonumber\\
 \silhc{W}&=-\myc{{WW}} \nonumber\\
 \silhc{B}&=-\myc{{BW}}+\myc{{WW}} \nonumber\\
 \silhc{{2 W}}&=2 g^2 \myc{{DW}} \nonumber\\
 \silhc{{2 B}}&=2 {g'}^2 \myc{{DB}} \nonumber\\
 \silhc{{BB}}&=-\frac{1}{4}\myc{{BB}}+\frac{1}{4}\myc{{BW}}-\frac{1}{4}\myc{{WW}} \nonumber\\
 \silhc{{HW}}&=\frac{1}{2}\myc{W}+\myc{{WW}} \nonumber\\
 \silhc{{HB}}&=\frac{1}{2}\myc{B}+\myc{{BW}}-\myc{{WW}} \nonumber\\
 \silhc{{3 W}}&=\frac{3}{2} g^2 \myc{{WWW}}+6 g^2 \myc{{DW}}
    \end{flalign}
  \item From our convention to SILH :
    \begin{flalign}
      \myc{{DW}}&=\frac{1}{2 g^2}\silhc{{2 W}} \nonumber\\
      \myc{{DB}}&=\frac{1}{2 {g'}^2}\silhc{{2 B}} \nonumber\\
 \myc{{BW}}&=-\silhc{B}-\silhc{W} \nonumber\\
 \myc{{\Phi,1}}&=-2 \silhc{T} \nonumber\\
 \myc{{\Phi,2}}&=\silhc{H}+\silhc{T} \nonumber\\
 \myc{{\Phi,3}}&=3 \lambda  \silhc{6} \nonumber\\
 \myc{{WWW}}&=\frac{2}{3 g^2} \silhc{{3 W}}-\frac{2}{g^2} \silhc{{2 W}} \nonumber\\
 \myc{{WW}}&=-\silhc{W} \nonumber\\
 \myc{{BB}}&=-\silhc{B}-4 \silhc{{BB}} \nonumber\\
 \myc{W}&=2 \silhc{W}+2 \silhc{{HW}} \nonumber\\
 \myc{B}&=2 \silhc{B}+2 \silhc{{HB}}
    \end{flalign}
\end{itemize}


\section*{Appendix B: Self Energies in $R_\xi$ Gauge }

In this appendix we present the detailed formulas of the self-energies that contribute to $\Pi_{XY}$ defined in Eq. 17 in $R_\xi$ gauge.
We use different gauge parameters $\xi_W$ and $\xi_Z$ for the W and Z bosons, respectively.
$\Pi_{XY}$ can be written as
\begin{equation}
\Pi_{XY}(q^2) = \Sigma_i f_i\biggl\{\Pi^i_{XY,FG}(q^2) + \Pi^i_{XY,\xi }(q^2)\biggr\},
\end{equation}
where XY represents $\gamma\gamma$, Z $\gamma$, ZZ and WW and $\Pi^i$ is the
part of the 2-point function which is proportional to $f_i$. $\Pi_{XY,FG}$ is 
independent of $\xi$ and contains 
results in Feynman gauge and the second term collects terms that vanish when $\xi = 1$. 
In the following, we express our results in terms of scalar integral functions $A_0$ and 
$B_0$ \cite{Passarino:1978jh}.
Only non-zero contributions are listed here.  We separate the contributions proportion to each $f_i$.


\allowdisplaybreaks
\noindent $\mathcal{O}_{B}$:
\begin{align}
\Pi_{\gamma \gamma,FG} \quad & = \quad -\frac{f_B}{\Lambda^2}\frac{1}{72 \pi} q^2 \alpha \bigl\{ 3 (20 m_W^2 +q^2) B_0(q^2,m_W^2,m_W^2)-6 A_0(m_W^2)+2(q^2-6 m_W^2)
           \bigr\}
\\
\Pi_{\gamma\gamma,\xi} \quad & = \quad \frac{f_B}{\Lambda^2}\frac{\alpha}{48 \pi  m_W^2} \left\{-2 q^2 \left(-2 q^2 m_W^2 \left(\xi_W-2\right)+m_W^4\left(\xi_W^2+4 \xi_W-5\right)+q^4\right) B_0\left(q^2,m_W^2,m_W^2\xi_W\right) 
\right. \nonumber \\ & \qquad \left. 
+\left(q^6-4 q^4 m_W^2 \xi_W\right) B_0\left(q^2,m_W^2 \
\xi_W,m_W^2 \xi_W\right)+\left(8 q^4 m_W^2+q^6\right) \
B_0\left(q^2,m_W^2,m_W^2\right) 
\right. \nonumber \\ & \qquad \left. 
-2 q^2 m_W^2 \left(\xi_W+5\right) \
A_0\left(m_W^2\right)+2 q^2 m_W^2 \left(\xi_W+5\right) \
A_0\left(m_W^2 \xi_W\right)\right\}
\\
\Pi_{\gamma Z,FG} \quad & = \quad \frac{f_B}{\Lambda^2}\frac{\alpha}{288 \pi  c s} \left\{3 \left(4 m_Z^2 \left(m_H^2+q^2\right)-2 q^2 
m_H^2+m_H^4-5 m_Z^4+q^4\right) B_0\left(q^2,m_H^2,m_Z^2\right) 
\right. \nonumber \\ & \qquad \left.
+3 q^2 \left(16 \left(5 s^2-2\right) m_W^2+q^2 \left(4 s^2-1\right)\right) \
B_0\left(q^2,m_W^2,m_W^2\right)+3 A_0\left(m_Z^2\right) 
\right. \nonumber \\ & \qquad \left.
\left(m_H^2+5 m_Z^2-q^2\right)-3 A_0\left(m_H^2\right) \left(m_H^2+5 \
m_Z^2+q^2\right)+6 q^2 \left(1-4 s^2\right) A_0\left(m_W^2\right)
\right.  \nonumber\\ & \qquad \left.
+2 q^2 \left(-3 m_H^2+\left(6-24 s^2\right) m_W^2-3 m_Z^2+4 q^2 \
s^2\right)\right\}
\\
\Pi_{\gamma Z,\xi} \quad & = \quad \frac{f_B}{\Lambda^2}\frac{\alpha}{96 \pi  c s m_W^2}  \left\{\left(m_W^2+q^2 \left(2 s^2-1\right)\right) \left(4 q^2 \
m_W^2 \xi_W-q^4\right) B_0\left(q^2,m_W^2 \xi_W,m_W^2 \xi_W\right) 
\right. \nonumber \\ & \qquad \left. 
+2 \left(m_W^2+q^2 \left(2 s^2-1\right)\right) \left(-2 q^2 m_W^2 \
\left(\xi_W-2\right)+m_W^4 \left(\xi_W^2+4 \xi_W-5\right)+q^4\right) 
\right. \nonumber \\ & \qquad \left.
\times B_0\left(q^2,m_W^2,m_W^2 \xi_W\right)-\left(8 q^2 m_W^2+q^4\right) \
\left(m_W^2+q^2 \left(2 s^2-1\right)\right) \
B_0\left(q^2,m_W^2,m_W^2\right) 
\right. \nonumber \\ & \qquad \left. 
+2 m_W^2 \left(\xi_W+5\right) \
A_0\left(m_W^2\right) \left(m_W^2+q^2 \left(2 s^2-1\right)\right)
\right. \nonumber \\ & \qquad \left. 
-2 m_W^2 \left(\xi_W+5\right) A_0\left(m_W^2 \xi_W\right) \
\left(m_W^2+q^2 \left(2 s^2-1\right)\right)\right\}
\\
\Pi_{ZZ,FG} \quad & = \quad \frac{f_B}{\Lambda^2}\frac{\alpha}{144 \pi  c^2 q^2}  \left\{
\right. \nonumber\\ & \quad \left.
\hspace{-4pt}\left(-3 m_H^4 \left(m_Z^2+q^2\right)+6 m_H^2 \left(-4q^2 m_Z^2+m_Z^4+q^4\right)
-3\left(m_Z^4-q^4\right)\left(m_Z^2-q^2\right)\right)B_0\hspace{-2pt}\left(q^2,m_H^2,m_Z^2\right)
\right. \nonumber\\ & \quad \left. 
+3 q^4 \
\left(q^2 \left(1-2 s^2\right)-8 \left(5 s^2-4\right) m_W^2\right) \
B_0\left(q^2,m_W^2,m_W^2\right) 
\right. \nonumber\\ & \quad \left. 
+3 A_0\left(m_H^2\right) \
\left(m_Z^2+q^2\right) \left(m_H^2-m_Z^2+q^2\right) 
\right. \nonumber\\ & \quad \left. 
+3 \
A_0\left(m_Z^2\right) \left(-m_H^2 \left(m_Z^2+q^2\right)-10 q^2 \
m_Z^2+m_Z^4+q^4\right) +6 q^4 \left(2 s^2-1\right) \
A_0\left(m_W^2\right) 
\right. \nonumber\\ & \quad \left. 
+2 q^2 \left(3 m_H^2 \left(m_Z^2+q^2\right) 
-2 q^2 \
\left(\left(3-6 s^2\right) m_W^2+q^2 s^2\right)+2 q^2 m_Z^2+3 \
m_Z^4\right)\right\}
\\
\Pi_{ZZ,\xi} \quad & = \quad \frac{f_B}{\Lambda^2}\frac{\alpha}{48 \pi  m_W^2}  \left\{\left(m_Z^2-q^2\right) \left(q^4-4 q^2 m_W^2 \xi_W\right) 
B_0\left(q^2,m_W^2 \xi_W,m_W^2 \xi_W\right) 
\right. \nonumber \\ & \qquad \left. 
-2 \left(m_Z^2-q^2\right) \left(-2 q^2 m_W^2 \left(\xi_W-2\right)+m_W^4 \
\left(\xi_W^2+4 \xi_W-5\right)+q^4\right) B_0\left(q^2,m_W^2,m_W^2 \
\xi_W\right) 
\right. \nonumber \\ & \qquad \left. 
+\left(8 q^2 m_W^2+q^4\right) \left(m_Z^2-q^2\right) \
B_0\left(q^2,m_W^2,m_W^2\right) 
\right. \nonumber \\ & \qquad \left. 
+2 m_W^2 \left(\xi_W+5\right) A_0\left(m_W^2\right) \left(q^2-m_Z^2\right)+2 m_W^2 \left\{\xi_W+5\right) \left(m_Z^2-q^2\right) 
A_0\left(m_W^2 \xi_W\right)\right\}
\\
\Pi_{WW,FG} \quad & = \quad \frac{f_B}{\Lambda^2}\frac{\alpha}{144 \pi  q^2}  \left\{12 \left(m_W^3-q^2 m_W\right){}^2 \
B_0\left(q^2,0,m_W^2\right) 
\right. \nonumber \\ & \qquad \left. 
-3 \left(-5 q^4 m_Z^2-m_W^4 \left(7 m_Z^2+8 \
q^2\right)+4 q^2 m_Z^4 
\right.\right. \nonumber \\ & \qquad \left. \left.
+2 m_W^2 \left(-14 q^2 m_Z^2+m_Z^4+2 q^4\right)+4 \
m_W^6+m_Z^6\right) B_0\left(q^2,m_W^2,m_Z^2\right) 
\right. \nonumber \\ & \qquad \left. 
-3 m_Z^2 A_0\left(m_W^2\right) \left(3 m_W^2+m_Z^2+5 q^2\right) 
\right. \nonumber \\ & \qquad \left. 
-3 A_0\left(m_Z^2\right) \left(-m_W^2 \left(3 m_Z^2+4 q^2\right)+5 q^2 \
m_Z^2+4 m_W^4-m_Z^4\right) 
\right. \nonumber \\ & \qquad \left. 
+2 q^2 m_Z^2 \left(-3 m_W^2+3 \
m_Z^2-q^2\right)\right\}
\\
\Pi_{WW,\xi} \quad & = \quad \frac{f_B}{\Lambda^2}\frac{\alpha}{48 \pi  q^2}  \left\{\left(m_W^2-q^2\right){}^2 \left(m_W^2+5 q^2\right) \
B_0\left(q^2,0,m_W^2\right) 
\right. \nonumber \\ & \qquad \left. 
+\left(q^2-m_W^2\right) \left(4 q^2 m_W^2 \
\xi_W+m_W^4 \xi_W^2-5 q^4\right) B_0\left(q^2,0,m_W^2 \xi_W\right) 
\right. \nonumber \\ & \qquad \left. 
+\left(m_W^2-q^2\right) \left(4 q^2 m_W^2 \xi_W+m_Z^2 \left(4 \
q^2-2 m_W^2 \xi_W\right)+m_W^4 \xi_W^2+m_Z^4-5 q^4\right) \
\right. \nonumber \\ & \qquad \left.
\times B_0\left(q^2,m_Z^2,m_W^2 \xi_W\right)+\left(q^2-m_W^2\right) \left(-2 \
m_W^2 \left(m_Z^2-2 q^2\right)+4 q^2 m_Z^2+m_W^4+m_Z^4-5 q^4\right) \
\right. \nonumber \\ & \qquad \left.
\times B_0\left(q^2,m_W^2,m_Z^2\right)+m_W^2 \left(\xi_W-1\right) \
A_0\left(m_Z^2\right) \left(m_W^2-q^2\right) 
\right. \nonumber \\ & \qquad \left. 
+m_Z^2 \
\left(m_W^2-q^2\right) A_0\left(m_W^2 \xi_W\right)+m_Z^2 \
A_0\left(m_W^2\right) \left(q^2-m_W^2\right)\right\}
\end{align}
\noindent $\mathcal{O}_{W}$:
\begin{align}
        \Pi_{\gamma \gamma,FG} \quad & = \quad -\frac{f_W}{\Lambda^2}\frac{1}{72 \pi} q^2 \alpha \left\{ 3 (20 m_W^2 +q^2) B0(q^2,m_W^2,m_W^2)-6 A0(m_W^2)+2(q^2-6 m_W^2)
           \right\}   
           \\
\Pi_{\gamma\gamma,\xi} \quad & = \quad \frac{f_W}{\Lambda^2}\frac{\alpha}{48 \pi  m_W^2}  \left\{-2 q^2 \left(-2 q^2 m_W^2 \left(\xi_W-2\right)+m_W^4\left(\xi_W^2+4 \xi_W-5\right)+q^4\right) B_0\left(q^2,m_W^2,m_W^2\xi_W\right) 
\right. \nonumber \\ & \qquad \left. 
+\left(q^6-4 q^4 m_W^2 \xi_W\right) B_0\left(q^2,m_W^2 \
\xi_W,m_W^2 \xi_W\right)+\left(8 q^4 m_W^2+q^6\right) \
B_0\left(q^2,m_W^2,m_W^2\right) 
\right. \nonumber \\ & \qquad \left. 
-2 q^2 m_W^2 \left(\xi_W+5\right) \
A_0\left(m_W^2\right)+2 q^2 m_W^2 \left(\xi_W+5\right) \
A_0\left(m_W^2 \xi_W\right)\right\}
\\           
\Pi_{\gamma Z,FG} \quad & = \quad \frac{f_W}{\Lambda^2}\frac{\alpha}{288 \pi  c s}  \left\{-3 \left(4 m_Z^2 \left(m_H^2+q^2\right)-2 q^2 \
m_H^2+m_H^4-5 m_Z^4+q^4\right) B_0\left(q^2,m_H^2,m_Z^2\right)
\right.  \nonumber\\ & \qquad \left. 
-3 \left(4 q^2 \left(21-20 s^2\right) m_W^2+48 m_W^4+q^4 \left(3-4 \
s^2\right)\right) B_0\left(q^2,m_W^2,m_W^2\right)
\right.  \nonumber\\ & \qquad \left.
-3 A_0\left(m_Z^2\right) \left(m_H^2+5 m_Z^2-q^2\right)+3 \
A_0\left(m_H^2\right) \left(m_H^2+5 m_Z^2+q^2\right)
\right.  \nonumber\\ & \qquad \left. 
+6 A_0\left(m_W^2\right) \left(24 m_W^2+q^2 \left(3-4 s^2\right)\right)
\right.  \nonumber\\ & \qquad \left. +2 \
\left(q^2 \left(3 m_H^2+3 m_Z^2+4 q^2
\left(s^2-1\right)\right)+6 q^2 \
\left(3-4 s^2\right) m_W^2-72 m_W^4 \right) \right\}
\\
\Pi_{\gamma Z,\xi} \quad & = \quad \frac{f_W}{\Lambda^2}\frac{\alpha}{96 \pi  s m_W m_Z}  \left\{q^2 \left(2 q^2-m_Z^2\right) \left(q^2-4 m_W^2 \xi_W\right) 
B_0\left(q^2,m_W^2 \xi_W,m_W^2 \xi_W\right) 
\right. \nonumber \\ & \qquad \left. 
+\left(4 m_W^4 \left(\xi_W-1\right) \left(m_Z^2 \left(\xi_W+2\right)-q^2 \left(\xi_W+5\right)\right) 
\right.\right. \nonumber \\ & \qquad \left.\left. +4 q^2 m_W^2 \left(m_Z^2 \left(\xi_W+4\right)+2 q^2 \
\left(\xi_W-2\right)\right)-4 \left(2 q^4 m_Z^2+q^6\right)\right) \
B_0\left(q^2,m_W^2,m_W^2 \xi_W\right) 
\right. \nonumber \\ & \qquad \left.
+\left(9 q^4 m_Z^2-8 m_W^2 \
\left(3 q^2 m_Z^2-2 q^4\right)+2 q^6\right) \
B_0\left(q^2,m_W^2,m_W^2\right) 
\right. \nonumber \\ & \qquad \left.
+A_0\left(m_W^2\right) \
\left(m_W^2 \left(4 m_Z^2 \left(\xi_W+2\right)-4 q^2 \left(\xi_W+5\right)\right)-10 q^2 m_Z^2\right)
\right. \nonumber \\ & \qquad \left.
+A_0\left(m_W^2 \xi_W\right) \
\left(m_W^2 \left(4 q^2 \left(\xi_W+5\right)-4 m_Z^2 \left(\xi_W+2\right)\right)+10 q^2 m_Z^2\right)
\right. \nonumber \\ & \qquad \left.
-4 q^2 m_W^2 m_Z^2 \left(\xi_W-1\right)\right\}
\\
\Pi_{ZZ,FG} \quad & = \quad -\frac{f_W}{\Lambda^2}\frac{\alpha}{144 \pi  q^2 s^2} \left\{
\right. \nonumber \\ & \qquad \left. 
3 \left(m_H^4 \left(m_Z^2+q^2\right)-2 m_H^2 \left(-4 \
q^2 m_Z^2+m_Z^4+q^4\right)+\left(m_Z^2-q^2\right)^2
\left(m_Z^2+q^2\right)\right) 
\right. \nonumber \\ & \qquad \left.
\times B_0\left(q^2,m_H^2,m_Z^2\right)+3 q^2 \
B_0\left(q^2,m_W^2,m_W^2\right) \left(4 m_W^2 \left(4 m_Z^2+q^2 \
\left(11-10 s^2\right)\right)
\right.\right. \nonumber \\ & \qquad \left.\left.
-10 q^2 m_Z^2+48 m_W^4+q^4 \left(1-2 \
s^2\right)\right)-3 A_0\left(m_H^2\right) \left(m_Z^2+q^2\right) \
\left(m_H^2-m_Z^2+q^2\right) 
\right. \nonumber \\ & \qquad \left.
+A_0\left(m_Z^2\right) \left(3 m_H^2 \
\left(m_Z^2+q^2\right)-3 \left(-10 q^2 m_Z^2+m_Z^4+q^4\right)\right)
\right. \nonumber \\ & \qquad \left.
-6 q^2 A_0\left(m_W^2\right) \left(24 m_W^2-10 m_Z^2+q^2 \left(1-2 \
s^2\right)\right)-2 q^2 \left(3 m_H^2 \left(m_Z^2+q^2\right)
\right.\right. \nonumber \\ & \qquad \left.\left.
+6 m_W^2 \left(2 \
m_Z^2+q^2 \left(1-2 s^2\right)\right)-72 m_W^4+3 m_Z^4+2 q^4 s^2-2 \
q^4\right)\right\}
\\
\Pi_{ZZ,\xi} \quad & = \quad -\frac{f_W}{\Lambda^2}\frac{\alpha}{48 \pi  q^2 s^2 m_Z^2}  \left\{\left(m_Z^2-q^2\right) \left(q^6-4 q^4 m_W^2 \xi_W\right) 
B_0\left(q^2,m_W^2 \xi_W,m_W^2 \xi_W\right) 
\right. \nonumber \\ & \qquad \left. 
-2 \left(m_Z^2-q^2\right) \left(5 q^4 m_Z^2-2 q^2 m_W^2 \left(2 m_Z^2 \left(\xi_W+1\right)+q^2 \left(\xi_W-2\right)\right)
\right.\right. \nonumber \\ & \qquad \left.\left.
+m_W^4 \left(\xi_W-1\right) \
\left(m_Z^2 \left(1-\xi_W\right)+q^2 \left(\xi_W+5\right)\right)+q^6\right) B_0\left(q^2,m_W^2,m_W^2 \xi_W\right)
\right. \nonumber \\ & \qquad \left.
+\left(m_Z^2-q^2\right) \left(10 q^4 m_Z^2+8 q^2 m_W^2 \left(q^2-2 \
m_Z^2\right)+q^6\right) B_0\left(q^2,m_W^2,m_W^2\right)
\right. \nonumber \\ & \qquad \left.
-2 A_0\left(m_W^2\right) \left(m_Z^2-q^2\right) \left(m_W^2 \left(m_Z^2 \
\left(1-\xi_W\right)+q^2 \left(\xi_W+5\right)\right)+5 q^2 m_Z^2\right)
\right. \nonumber \\ & \qquad \left.
+2 \left(m_Z^2-q^2\right) A_0\left(m_W^2 \xi_W\right) \left(m_W^2 \
\left(m_Z^2 \left(1-\xi_W\right)+q^2 \left(\xi_W+5\right)\right)+5 q^2 \
m_Z^2\right)
\right. \nonumber \\ & \qquad \left.
+4 q^2 m_W^2 m_Z^2 \left(\xi_W-1\right) \
\left(q^2-m_Z^2\right)\right\}
\\
\Pi_{WW,FG} \quad & = \quad -\frac{f_W}{\Lambda^2}\frac{\alpha}{144 \pi  q^2 s^2}  \left\{
\right. \nonumber \\ & \qquad \left.
3 \left(m_H^4 \left(m_W^2+q^2\right)-2 m_H^2 \left(-4 \
q^2 m_W^2+m_W^4+q^4\right)+\left(m_W^2-q^2\right){}^2 \
\left(m_W^2+q^2\right)\right) 
\right. \nonumber \\ & \qquad \left.
\times B_0\left(q^2,m_H^2,m_W^2\right)-12 s^2 \
m_W^2 \left(m_W^2-q^2\right){}^2 B_0\left(q^2,0,m_W^2\right)
\right. \nonumber \\ & \qquad \left.
+3 \left(4 \
q^4 m_Z^2+m_W^4 \left(8 m_Z^2+q^2 \left(59-8 s^2\right)\right)-5 q^2 m_Z^4
\right.\right. \nonumber \\ & \qquad \left.\left.
+m_W^2 \left(10 q^2 m_Z^2-6 m_Z^4+2 q^4 \left(2 \
s^2+15\right)\right)+\left(4 s^2-2\right) m_W^6+q^6\right) \
B_0\left(q^2,m_W^2,m_Z^2\right) 
\right. \nonumber \\ & \qquad \left.
-3 A_0\left(m_H^2\right) \
\left(m_W^2+q^2\right) \left(m_H^2-m_W^2+q^2\right)
\right. \nonumber \\ & \qquad \left.
+3 \
A_0\left(m_W^2\right) \left(m_H^2 \left(m_W^2+q^2\right)+m_W^2 \
\left(q^2-6 m_Z^2\right)-5 q^2 m_Z^2+5 m_W^4-2 q^4\right)
\right. \nonumber \\ & \qquad \left.
+3 A_0\left(m_Z^2\right) \left(m_W^2 \left(6 m_Z^2-q^2 \left(4 \
s^2+19\right)\right)+5 q^2 m_Z^2+\left(4 s^2-6\right) m_W^4-q^4\right)
\right. \nonumber \\ & \qquad \left.
+2 q^2 \
\left(-3 m_H^2 \left(m_W^2+q^2\right)+m_W^2 \left(36 m_Z^2-3 q^2\right)-3 \
q^2 m_Z^2+21 m_W^4+2 q^4\right)\right\}
\\
\Pi_{WW,\xi} \quad & = \quad -\frac{f_W}{\Lambda^2}\frac{\alpha}{48 \pi  q^2 s^2 m_Z^2}  \left\{
\right. \nonumber \\ & \qquad \left. 
\left(m_W^2-q^2\right){}^2 \left(2 m_W^2 \left(5 \
q^2-m_Z^2 \xi_Z\right)+\left(q^2-m_Z^2 \xi_Z\right){}^2+m_W^4\right) \
B_0\left(q^2,m_W^2,m_Z^2 \xi_Z\right)
\right. \nonumber \\ & \qquad \left.
-q^2 \left(q^2-m_W^2\right) \
\left(-2 q^2 \left(m_W^2 \xi_W+m_Z^2 \xi_Z\right)+\left(m_W^2 \xi_W-m_Z^2 \
\xi_Z\right){}^2+q^4\right) 
\right. \nonumber \\ & \qquad \left.
\times B_0\left(q^2,m_W^2 \xi_W,m_Z^2 \xi_Z\right)
\right. \nonumber \\ & \qquad \left.
-\left(m_W^2-q^2\right) \left(m_W^2-m_Z^2\right) \left(4 q^2 m_W^2 \
\xi_W+m_W^4 \xi_W^2-5 q^4\right) B_0\left(q^2,0,m_W^2 \xi_W\right)
\right. \nonumber \\ & \qquad \left.
+\left(m_W^2-q^2\right) \left(4 q^2 m_W^2+m_W^4-5 q^4\right) \
\left(m_W^2-m_Z^2\right) \
B_0\left(q^2,0,m_W^2\right)
\right. \nonumber \\ & \qquad \left.
+\left(m_W^2-q^2\right) \left(-4 q^4 \
m_Z^2-m_W^4 \xi_W \left(2 m_Z^2+q^2 \left(\xi_W-4\right)\right)+5 q^2 m_Z^4
\right.\right. \nonumber \\ & \qquad \left.\left.
+m_W^2 \left(-4 q^2 m_Z^2 \left(\xi_W-1\right)+m_Z^4+q^4 \left(2 \xi_W-5\right)\right)+m_W^6 \xi_W^2-q^6\right) B_0\left(q^2,m_Z^2,m_W^2 \
\xi_W\right)
\right. \nonumber \\ & \qquad \left.
-\left(q^2-m_W^2\right) \left(4 q^4 m_Z^2+m_W^4 \left(4 \
m_Z^2-13 q^2\right)-5 q^2 m_Z^4-2 m_W^2 \left(-q^2 m_Z^2+m_Z^4-7 q^4\right)
\right.\right. \nonumber \\ & \qquad \left.\left.
-2 m_W^6+q^6\right) B_0\left(q^2,m_W^2,m_Z^2\right)
\right. \nonumber \\ & \qquad \left.
+m_W^2 \
A_0\left(m_Z^2\right) \left(m_W^2-q^2\right) \left(m_W^2 \left(\xi_W-2\right)+m_Z^2+q^2 \left(-\left(\xi_W+10\right)\right)\right)
\right. \nonumber \\ & \qquad \left.
+m_W^2 \
\left(m_W^2-q^2\right) A_0\left(m_Z^2 \xi_Z\right) \left(m_W^2-m_Z^2 \
\xi_Z+q^2 \left(\xi_W+10\right)\right)
\right. \nonumber \\ & \qquad \left.
-m_Z^2 \left(q^2-m_W^2\right) \
A_0\left(m_W^2 \xi_W\right) \left(m_W^2 \left(1-\xi_W\right)+q^2 \
\left(\xi_Z+10\right)\right)
\right. \nonumber \\ & \qquad \left.
+m_Z^2 A_0\left(m_W^2\right) \
\left(q^2-m_W^2\right) \left(m_W^2 \left(1-\xi_Z\right)+q^2 \left(\xi_Z+10\right)\right)
\right. \nonumber \\ & \qquad \left.
+2 q^2 m_W^2 m_Z^2 \left(q^2-m_W^2\right) \left(\xi_W+\xi_Z-2\right)\right\}
\end{align}  

\noindent $\mathcal{O}_{BB}$:
\begin{align}
\Pi_{\gamma \gamma,FG} \quad & = \quad -\frac{f_{BB}}{\Lambda^2}\frac{\alpha}{4 \pi  m_H^2}  q^2 \left\{m_H^2 A_0\left(m_H^2\right)+6 m_W^2 
                        A_0\left(m_W^2\right)+3 m_Z^2 A_0\left(m_Z^2\right) \right. \\
                                &  \qquad \left. -2\left(2 m_W^4+m_Z^4\right)\right\}
\\
\Pi_{\gamma Z,FG} \quad & = \quad \frac{f_{BB}}{\Lambda^2}\frac{\alpha}{4 \pi  \
c s m_H^2}  \left(c^2-1\right) \left\{m_H^2 m_Z^2 \
\left(-m_H^2+m_Z^2+q^2\right) B_0\left(q^2,m_H^2,m_Z^2\right) 
\right.  \nonumber\\ & \qquad \left.
-m_Z^2 \
A_0\left(m_Z^2\right) \left(m_H^2+3 q^2\right)+m_H^2 \
A_0\left(m_H^2\right) \left(m_Z^2-q^2\right)-6 q^2 m_W^2 \
A_0\left(m_W^2\right) 
\right.  \nonumber\\ & \qquad \left.
+2 q^2 \left(2 m_W^4+m_Z^4\right)\right\}
\\            
\Pi_{ZZ,FG} \quad & = \quad \frac{f_{BB}}{\Lambda^2}\frac{\alpha}{4 \pi  c^2 m_H^2}  \left(1-c^2\right) \left\{2 m_H^2 m_Z^2 \
\left(-m_H^2+m_Z^2+q^2\right) B_0\left(q^2,m_H^2,m_Z^2\right)
\right. \nonumber \\ & \qquad \left. 
+m_H^2 A_0\left(m_H^2\right) \left(2 m_Z^2-q^2\right)-m_Z^2 \
A_0\left(m_Z^2\right) \left(2 m_H^2+3 \left(m_Z^2+q^2\right)\right)
\right. \nonumber \\ & \qquad \left. 
-6 q^2 m_W^2 A_0\left(m_W^2\right)+2 \left(2 q^2 m_W^4+q^2 \
m_Z^4+m_Z^6\right)\right\}
\\
\Pi_{WW,FG} \quad & = \quad -\frac{f_{BB}}{\Lambda^2}\frac{\alpha}{4 \pi  m_H^2}  \left(c^2-1\right) m_Z^4 \left\{2 m_Z^2-3 \
A_0\left(m_Z^2\right)\right\}
\end{align}

\noindent $\mathcal{O}_{WW}$:
\begin{align}
\Pi_{\gamma \gamma,FG} \quad & = \quad -\frac{f_{WW}}{\Lambda^2}\frac{\alpha}{4 \pi  m_H^2} q^2 \left\{m_H^2 A_0\left(m_H^2\right)+6 m_W^2 \
A_0\left(m_W^2\right)
\right.  \nonumber\\ & \qquad \left.
+3 m_Z^2 A_0\left(m_Z^2\right)-2 \left(2 \
m_W^4+m_Z^4\right)\right\}
\\
\Pi_{\gamma Z,FG} \quad & = \quad \frac{f_{WW}}{\Lambda^2}\frac{\alpha}{4 \pi  s m_H^2} c \left\{m_H^2 m_Z^2 \left(-m_H^2+m_Z^2+q^2\right) \
B_0\left(q^2,m_H^2,m_Z^2\right)
\right.  \nonumber\\ & \qquad \left.
-m_Z^2 A_0\left(m_Z^2\right) 
\left(m_H^2+3 q^2\right)   
+m_H^2 A_0\left(m_H^2\right) 
\left(m_Z^2-q^2\right)
\right.  \nonumber\\ & \qquad \left.
-6 q^2 m_W^2 A_0\left(m_W^2\right)+2 q^2 \left(2 \
m_W^4+m_Z^4\right)\right\}
\\
\Pi_{ZZ,FG} \quad & = \frac{f_{WW}}{\Lambda^2}\frac{\alpha}{4 \pi  s^2 m_H^2}  c^2 \left\{2 m_H^2 m_Z^2 \left(-m_H^2+m_Z^2+q^2\right) \
B_0\left(q^2,m_H^2,m_Z^2\right)
\right.  \nonumber\\ & \qquad \left.
+m_H^2 A_0\left(m_H^2\right) \
\left(2 m_Z^2-q^2\right)-m_Z^2 A_0\left(m_Z^2\right) \left(2 m_H^2+3 \
\left(m_Z^2+q^2\right)\right)
\right.  \nonumber\\ & \qquad \left.
-6 A_0\left(m_W^2\right) \left(q^2 \
m_W^2+m_Z^4\right)+2 \left(2 q^2 m_W^4+q^2 m_Z^4+2 m_W^2 \
m_Z^4+m_Z^6\right)\right\}
\\
\Pi_{WW,FG} \quad & = \quad -\frac{f_{WW}}{\Lambda^2}\frac{\alpha}{4 \pi  s^2 m_H^2}  \left\{-2 m_H^2 m_W^2 \left(-m_H^2+m_W^2+q^2\right) \
B_0\left(q^2,m_H^2,m_W^2\right)
\right.  \nonumber\\ & \qquad \left.
+m_H^2 A_0\left(m_H^2\right) \
\left(q^2-2 m_W^2\right)+2 m_W^2 A_0\left(m_W^2\right) \left(m_H^2+3 \
\left(m_W^2+q^2\right)\right)
\right.  \nonumber\\ & \qquad \left.
+3 A_0\left(m_Z^2\right) \left(q^2 \
m_Z^2+m_W^4\right)-2 \left(m_W^4 \left(m_Z^2+2 q^2\right)+q^2 m_Z^4+2 \
m_W^6\right)\right\}
\end{align}

\noindent $\mathcal{O}_{WWW}$:
\begin{align}
\Pi_{\gamma \gamma,FG} \quad & = \quad \frac{f_{WWW}}{\Lambda^2}\frac{q^2 s^2 g_w^4}{16 \pi ^2} \left\{3 \left(q^2-4 m_W^2\right) \
B_0\left(q^2,m_W^2,m_W^2\right)-6 A_0\left(m_W^2\right)+6 \
m_W^2-q^2\right\} \\
\Pi_{\gamma Z,FG} \quad & = \quad -\frac{f_{WWW}}{\Lambda^2}\frac{c q^2 s g_w^4}{16 \pi ^2} \left\{\left(12 m_W^2-3 q^2\right) \
B_0\left(q^2,m_W^2,m_W^2\right)+6 A_0\left(m_W^2\right)-6 \
m_W^2+q^2\right\}
\\
\Pi_{ZZ,FG} \quad & =-\frac{f_{WWW}}{\Lambda^2}\frac{c^2 q^2 g_w^4}{16 \pi ^2} \left\{\left(12 m_W^2-3 q^2\right) \
B_0\left(q^2,m_W^2,m_W^2\right)+6 A_0\left(m_W^2\right)-6 \
m_W^2+q^2\right\}
\\
\Pi_{WW,FG} \quad & = \quad -\frac{f_{WWW}}{\Lambda^2}\frac{g_w^4}{16 \pi ^2} \left\{-3 s^2 \left(m_W^2-q^2\right){}^2 \
B_0\left(q^2,0,m_W^2\right)+3 \left(-m_W^2 \left(m_Z^2+2 q^2 \
\left(s^2-2\right)\right)
\right.\right.  \nonumber\\ & \qquad \left.\left.
+\left(s^2+1\right) m_W^4+q^4 \
\left(s^2-1\right)\right) B_0\left(q^2,m_W^2,m_Z^2\right)
\right.  \nonumber\\ & \qquad \left.
+3 A_0\left(m_Z^2\right) \left(s^2 m_W^2-q^2 \left(s^2-1\right)\right)+3 \
q^2 A_0\left(m_W^2\right)-6 q^2 m_W^2+q^4\right\}
\end{align}

\noindent $\mathcal{O}_{\Phi,2}$:\footnote{The contributions to  $\Pi_{ZZ}$ and $\Pi_{WW}$ from ${\cal O}_{\Phi,2}$ are
gauge independent without the addition of tadpole diagrams.  Since the
tadpole diagrams cancel in the calculation of the oblique parameters, we
do not include them for the ${\cal O}_{\Phi,2}$ 2-point functions.  The tadpole
contributions $\it{are}$ included in the 2-point functions listed above for all other operators.}
\begin{align}
  \Pi_{ZZ}\quad & = \quad
  \frac{f_{\Phi,2}}{\Lambda^2}\frac{m_Z^2}{144\pi^2q^2}
  \left[ 3\left( m_H^4-2m_H^2(m_Z^2+q^2)+m_Z^4+10m_Z^2q^2+q^4 \right)
    B_0\left( q^2,m_H^2,m_Z^2 \right)
    \right.
    \nonumber\\ &
    \left.-3(m_H^2-m_Z^2-2q^2)A_0(m_H^2)+3(m_H^2-m_Z^2-q^2)A_0(m_Z^2)
    -2q^2(3m_H^2+3m_Z^2-q^2) \right]
    \\
  \Pi_{WW}\quad & = \quad
  \frac{f_{\Phi,2}}{\Lambda^2}\frac{m_W^2}{144\pi^2q^2}
  \left[ 3\left( m_H^4-2m_H^2(m_W^2+q^2)+m_W^4+10m_W^2q^2+q^4 \right)
    B_0\left( q^2,m_H^2,m_W^2 \right)
    \right.
    \nonumber\\ &
    \left.-3(m_H^2-m_W^2-2q^2)A_0(m_H^2)+3(m_H^2-m_W^2-q^2)A_0(m_W^2)
    -2q^2(3m_H^2+3m_W^2-q^2) \right]
\end{align}

\section*{Appendix C: Pinch Terms in $R_\xi$ Gauge}
Analytic results for
$\Gamma_L^V$ as defined in Eq. \ref{pinchdef} are given here in $R_\xi$ gauge and contain both a pinch
contribution and a non-pinch contribution,
\begin{equation}
\Gamma_L^V=\Gamma_L^V(pinch)+\Gamma_L^V(non-pinch)\, .
\end{equation}
The "pinch" contribution is defined as in Refs. \cite{Binosi:2009qm,Dawson:2007yk}, to be the contribution of the 3-point interaction which
exactly cancels the internal fermion propagator.  
We note that $\Gamma_L^V(non-pinch)$ is gauge invariant.  Ref. \cite{Mebane:2013zga,Mebane:2013cra} included this non-pinch contribution in
their definition of ${\overline \Pi}_{XY}$. 
From here on, we use $\Gamma_L^V$ to denote the pinch contribution only, as is used in the
definition of the oblique parameters in Ref. \cite{Degrassi:1993kn}.
We use different gauge parameters $\xi_W$ and $\xi_Z$ for the W and Z bosons, respectively, and they
cancel separately in our final result.  
$\Delta \Gamma_L^V$ can be written as
\begin{eqnarray}
\Delta \Gamma_L^V &= \Delta \Gamma_{L,FG}^V+ \Delta \Gamma_{L,\xi}^V, 
\end{eqnarray}
where V represents $\gamma$, Z and W. The first term $\Pi_{XY,FG}$ contains results in Feynman gauge and is 
independent of $\xi$. The second term collects terms that vanish when $\xi = 1$. 
In the following, we express our results in terms of scalar integral functions $A_0$ and $B_0$ \cite{Passarino:1978jh}. Only non-zero pinch contributions are listed here, and we separate the contributions in terms proportional to each of 
the individual $f_i$. 

\noindent $\mathcal{O}_{B}$:
\begin{align}
\Delta \Gamma_{L,\xi}^{\gamma} \quad & = \quad \frac{f_{B}}{\Lambda^2}\frac{\alpha}{96 \pi  s m_W^2}  \left\{
\right.  \nonumber\\ & \qquad \left.
q^2 \left(8 m_W^2+q^2\right) \
\left(-B_0\left(q^2,m_W^2,m_W^2\right)\right)+\left(4 q^2 m_W^2 \xi_W-q^4\right) B_0\left(q^2,m_W^2 \xi_W,m_W^2 \xi_W\right)
\right.  \nonumber\\ & \qquad \left.
+2 \left(-2 \
q^2 m_W^2 \left(\xi_W-2\right)+m_W^4 \left(\xi_W^2+4 \xi_W-5\right)+q^4\right) B_0\left(q^2,m_W^2,m_W^2 \xi_W\right)
\right.  \nonumber\\ & \qquad \left.
+2 m_W^2 \
\left(\xi_W+5\right) A_0\left(m_W^2\right)-2 m_W^2 \left(\xi_W+5\right) A_0\left(m_W^2 \xi_W\right)\right\}
\\
\Delta \Gamma_{L,\xi}^{Z} \quad & = \quad \frac{f_{B}}{\Lambda^2}\frac{\alpha}{96 \pi  c m_W^2}  \left\{\left(q^4-4 q^2 m_W^2 \xi_W\right) \
B_0\left(q^2,m_W^2 \xi_W,m_W^2 \xi_W\right)
\right.  \nonumber\\ & \qquad \left.
+\left(4 q^2 m_W^2 \
\left(\xi_W-2\right)-2 m_W^4 \left(\xi_W^2+4 \xi_W-5\right)-2 q^4\right) \
B_0\left(q^2,m_W^2,m_W^2 \xi_W\right)
\right.  \nonumber\\ & \qquad \left.
+\left(8 q^2 m_W^2+q^4\right) \
B_0\left(q^2,m_W^2,m_W^2\right)-2 m_W^2 \left(\xi_W+5\right) \
A_0\left(m_W^2\right)
\right.  \nonumber\\ & \qquad \left.
+2 m_W^2 \left(\xi_W+5\right) \
A_0\left(m_W^2 \xi_W\right)\right\}
\\
\Delta \Gamma_{L,FG}^{W} \quad & = \quad -\frac{f_{B}}{\Lambda^2}\frac{\alpha}{8 \pi }  c^2 m_Z^2 \
\left(B_0\left(q^2,0,m_W^2\right)-B_0\left(q^2,m_W^2,m_Z^2\right)\right)
\\
\Delta \Gamma_{L,\xi}^{W} \quad & = \quad \frac{f_{B}}{\Lambda^2}\frac{\alpha}{96 \pi  q^2}  \left\{\left(-4 q^2 m_W^2 \xi_W-m_W^4 \xi_W^2+5 q^4\right) \
B_0\left(q^2,0,m_W^2 \xi_W\right)
\right.  \nonumber\\ & \qquad \left.
+\left(4 q^2 m_W^2 \xi_W+m_Z^2 \
\left(4 q^2-2 m_W^2 \xi_W\right)+m_W^4 \xi_W^2+m_Z^4-5 q^4\right) \
B_0\left(q^2,m_Z^2,m_W^2 \xi_W\right)
\right.  \nonumber\\ & \qquad \left.
+\left(-4 q^2 m_W^2-4 q^2 \
m_Z^2+2 m_W^2 m_Z^2-m_W^4-m_Z^4+5 q^4\right) \
B_0\left(q^2,m_W^2,m_Z^2\right)
\right.  \nonumber\\ & \qquad \left.
+\left(4 q^2 m_W^2+m_W^4-5 q^4\right) \
B_0\left(q^2,0,m_W^2\right)+m_W^2 \left(\xi_W-1\right) \
A_0\left(m_Z^2\right)
\right.  \nonumber\\ & \qquad \left.
+m_Z^2 A_0\left(m_W^2 \xi_W\right)-m_Z^2 \
A_0\left(m_W^2\right)\right\}
\end{align}

\noindent $\mathcal{O}_{W}$:
\begin{align}
\Delta \Gamma_{L,\xi}^{\gamma} \quad & = \quad \frac{f_{W}}{\Lambda^2}\frac{\alpha}{96 \pi  s m_W^2}  \left\{q^2 \left(8 m_W^2+q^2\right) \
\left(-B_0\left(q^2,m_W^2,m_W^2\right)\right\}
\right. \nonumber \\ & \qquad \left. 
+\left(4 q^2 m_W^2 \xi_W-q^4\right) B_0\left(q^2,m_W^2 \xi_W,m_W^2 \xi_W\right)
\right.  \nonumber\\ & \qquad \left.
+2 \left(-2 \
q^2 m_W^2 \left(\xi_W-2\right)+m_W^4 \left(\xi_W^2+4 \xi_W-5\right)+q^4\right) B_0\left(q^2,m_W^2,m_W^2 \xi_W\right)
\right.  \nonumber\\ & \qquad \left.
+2 m_W^2 \
\left(\xi_W+5\right) A_0\left(m_W^2\right)-2 m_W^2 \left(\xi_W+5\right) A_0\left(m_W^2 \xi_W\right)\right)
\\
\Delta \Gamma_{L,FG}^{Z} \quad & = \quad -\frac{f_{W}}{\Lambda^2}\frac{\alpha}{4 \pi  s^2}  c m_Z^2 B_0\left(q^2,m_W^2,m_W^2\right)
\\
\Delta \Gamma_{L,\xi}^{Z} \quad & = \quad \frac{f_{W}}{\Lambda^2}\frac{\alpha}{96 \pi  q^2 m_W \left(c \ m_W-m_Z\right)}  \left\{\left(q^6-4 q^4 m_W^2 \xi_W\right) \
B_0\left(q^2,m_W^2 \xi_W,m_W^2 \xi_W\right)
\right.  \nonumber\\ & \qquad \left.
+\left(-2 m_W^4 \left(\xi_W-1\right) \left(m_Z^2 \left(1-\xi_W\right)+q^2 \left(\xi_W+5\right)\right)
\right.\right.  \nonumber\\ & \qquad \left.\left.
+4 q^2 m_W^2 \left(2 m_Z^2 \left(\xi_W+1\right)+q^2 \
\left(\xi_W-2\right)\right)-2 \left(5 q^4 m_Z^2+q^6\right)\right) \
B_0\left(q^2,m_W^2,m_W^2 \xi_W\right)
\right.  \nonumber\\ & \qquad \left.
+\left(10 q^4 m_Z^2+8 q^2 m_W^2 \
\left(q^2-2 m_Z^2\right)+q^6\right) \
B_0\left(q^2,m_W^2,m_W^2\right)
\right.  \nonumber\\ & \qquad \left.
+A_0\left(m_W^2\right) \left(-2 \
m_W^2 \left(m_Z^2 \left(1-\xi_W\right)+q^2 \left(\xi_W+5\right)\right)-10 \
q^2 m_Z^2\right)
\right.  \nonumber\\ & \qquad \left.
+A_0\left(m_W^2 \xi_W\right) \left(2 m_W^2 \
\left(m_Z^2 \left(1-\xi_W\right)+q^2 \left(\xi_W+5\right)\right)+10 q^2 \
m_Z^2\right)-4 q^2 m_W^2 m_Z^2 \left(\xi_W-1\right)\right\}
\\
\Delta \Gamma_{L,FG}^{W} \quad & = \quad \frac{f_{W}}{\Lambda^2}\frac{\alpha}{8 \pi  s^2}  c^2 m_Z^2 \left\{\left(s^2-2\right) \
B_0\left(q^2,m_W^2,m_Z^2\right)-s^2 \
B_0\left(q^2,0,m_W^2\right)\right\}
\\
\Delta \Gamma_{L,\xi}^{W} \quad & = \quad -\frac{f_{W}}{\Lambda^2}\frac{\alpha}{96 \pi  q^2 s^2 m_Z^2}  \left\{
\right.  \nonumber\\ & \qquad \left.
\left(m_W^2-q^2\right) \left(2 m_W^2 \left(5 q^2-m_Z^2 \
\xi_Z\right)+\left(q^2-m_Z^2 \xi_Z\right){}^2+m_W^4\right) \
B_0\left(q^2,m_W^2,m_Z^2 \xi_Z\right)
\right.  \nonumber\\ & \qquad \left.
+q^2 \left(-2 q^2 \left(m_W^2 \
\xi_W+m_Z^2 \xi_Z\right)+\left(m_W^2 \xi_W-m_Z^2 \xi_Z\right){}^2+q^4\right) B_0\left(q^2,m_W^2 \xi_W,m_Z^2 \xi_Z\right)
\right.  \nonumber\\ & \qquad \left.
-\left(m_W^2-m_Z^2\right) \left(4 q^2 m_W^2 \xi_W+m_W^4 \xi_W^2-5 \
q^4\right) B_0\left(q^2,0,m_W^2 \xi_W\right) 
\right.  \nonumber\\ & \qquad \left.
+\left(4 q^2 \
m_W^2+m_W^4-5 q^4\right) \left(m_W^2-m_Z^2\right) \
B_0\left(q^2,0,m_W^2\right)
\right.  \nonumber\\ & \qquad \left.
+\left(-4 q^4 m_Z^2-m_W^4 \xi_W \left(2 \
m_Z^2+q^2 \left(\xi_W-4\right)\right)+5 q^2 m_Z^4
\right.\right.  \nonumber\\ & \qquad \left.\left.
+m_W^2 \left(-4 q^2 m_Z^2 \
\left(\xi_W-1\right)+m_Z^4+q^4 \left(2 \xi_W-5\right)\right)+m_W^6 \xi_W^2-q^6\right) B_0\left(q^2,m_Z^2,m_W^2 \xi_W\right)
\right.  \nonumber\\ & \qquad \left.
+\left(4 q^4 \
m_Z^2+m_W^4 \left(4 m_Z^2-13 q^2\right)-5 q^2 m_Z^4-2 m_W^2 \left(-q^2 \
m_Z^2+m_Z^4-7 q^4\right)-2 m_W^6+q^6\right)
\right.  \nonumber\\ & \qquad \left.
\times B_0\left(q^2,m_W^2,m_Z^2\right)+m_W^2 A_0\left(m_Z^2\right) \
\left(m_W^2 \left(\xi_W-2\right)+m_Z^2+q^2 \left(-\left(\xi_W+10\right)\right)\right)
\right.  \nonumber\\ & \qquad \left.
+m_W^2 A_0\left(m_Z^2 \xi_Z\right) \
\left(m_W^2-m_Z^2 \xi_Z+q^2 \left(\xi_W+10\right)\right)
\right.  \nonumber\\ & \qquad \left.
+m_Z^2 \
A_0\left(m_W^2\right) \left(m_W^2 \left(\xi_Z-1\right)-q^2 \left(\xi_Z+10\right)\right)
\right.  \nonumber\\ & \qquad \left.
+m_Z^2 A_0\left(m_W^2 \xi_W\right) \left(m_W^2 \
\left(1-\xi_W\right)+q^2 \left(\xi_Z+10\right)\right)-2 q^2 m_W^2 m_Z^2 \
\left(\xi_W+\xi_Z-2\right)\right\}
\end{align}

\noindent $\mathcal{O}_{WWW}$:
\begin{align}
\Delta \Gamma_{L,FG}^{\gamma} \quad & = \quad -\frac{f_{WWW}}{\Lambda^2}\frac{3 \alpha}{8 \pi  s}  q^2 g_w^2 B_0\left(q^2,m_W^2,m_W^2\right)
\\
\Delta \Gamma_{L,FG}^{Z} \quad & = \quad -\frac{f_{WWW}}{\Lambda^2}\frac{3 \alpha}{8 \pi s^2}  c \ q^2 g_w^2 B_0\left(q^2,m_W^2,m_W^2\right)
\\ 
\Delta \Gamma_{L,FG}^{W} \quad & = \quad \frac{f_{WWW}}{\Lambda^2}\frac{-3 \alpha}{8 \pi  s^2}  q^2 g_w^2 \left(c^2 \
B_0\left(q^2,m_W^2,m_Z^2\right)+s^2 \
B_0\left(q^2,0,m_W^2\right)\right)
\end{align}

\section*{Appendix D: Analytic Results for Oblique Parameters}
The finite contributions to the oblique parameters are,
\newcommand{\rh}{\left(\frac{m_H}{m_Z}\right)}
\begin{flalign}
 \quad R_{S1}=& \;\;
  \frac{m_Z^2}{24\pi\Lambda^2}
  \left\{
    \frac{29}{3}\left( 5c^2+1 \right)f_B-c^2(64c^2+15)f_W
    +72\left(s^2f_{BB}+c^2f_{WW}\right)+72c^2f_{WWW}g^2
  \right.
  \nonumber\\&
  \left.
  +3\rh^2\left( c^2f_B+s^2f_W-8c^2f_{WW}-8s^2f_{BB} \right)
  +\rh^4\left[ \left( 2c^2-3\right)f_B+\left(2s^2-3\right)f_W \right]
  \right.
  \nonumber\\&
  \left.
  +\bigg[
   -48\left(s^2f_{BB}+c^2f_{WW}\right)
   +2\rh^2\left[(2s^2+1)f_B+(2c^2+1)f_W+12\left(s^2f_{BB}+c^2f_{WW}\right)\right]
   \right.
   \nonumber\\&
   \left.
   +\rh^4\left[(2s^2+1)f_B+(2c^2+1)f_W\right]
    \bigg]\rh\sqrt{4-\rh^2}\cos^{-1}\left(\frac{m_H}{2m_Z}\right)
  \right.
  \nonumber\\&
  \left.
  -2\left[\left( 32c^2+1 \right)f_B+36c^2f_{WWW}g^2-\left( 32c^4+8c^2-1 \right)f_W
    \right]\sqrt{4c^2-1}\sin^{-1}\left( \frac{1}{2c} \right)
  \right\} \nonumber
  \\
  &+\frac{f_{\Phi,2}}{\Lambda^2}\frac{m_W^2 s^2}{72\pi^2\alpha}\left[
    79-27\rh^2+6\rh^4
    \right.\nonumber\\&\left.
    -6\left(12-4\rh^2+\rh^4\right)\rh\sqrt{4-\rh^2}\cos^{-1}
    \left( \frac{m_H}{2m_Z} \right)\right]
\end{flalign}
\vspace{-20 pt}
\begin{eqnarray}
R_{S2}&=& {m_Z^2\over 12\pi\Lambda^2}\{
-(1+30c^2)f_B-(1-10c^2)f_W-36c^2 g^2 f_{WWW}\}\nonumber 
\\
R_{S3}&=&{m_H^2\over 24 \pi\Lambda^2}
\biggl\{(f_B+f_W)\biggl({m_H^4\over m_Z^4}-12\biggr)
+2(s^2 f_B+c^2 f_W)
\biggl({m_H^4\over m_Z^4}+12-18{m_H^2\over m_H^2-m_Z^2}\biggr) \nonumber \\&&
+24 (s^2 f_{BB}+c^2 f_{WW}) \biggl({m_H^2-4m_Z^2 \over m_Z^2}\biggr)
\biggr\}
-\frac{f_{\Phi,2}}{\Lambda^2}\frac{m_Z^4s^2c^2}{12\pi^2\alpha(m_H^2-m_Z^2)}
\nonumber\\&&\times
\rh^2\left[\rh^6-7\rh^4+24\rh^2-36\right]
\nonumber \\
R_{T1}&=&{m_Z^2\over 32 \pi 
\Lambda^2}\biggl\{15f_B+\biggl({3-23c^2\over s^2}\biggr)f_W\biggr\}+{5m_H^2\over 32\pi c^2\Lambda^2}f_B
-\frac{5m_Z^2s^2}{32\pi^2\alpha\Lambda^2}f_{\Phi,2}
\nonumber \\
R_{T2}&=& {m_W^2\over 8 \pi s^2\Lambda^2}\biggl\{ 5 f_B-3\biggl({m_W^2
\over m_H^2-m_W^2}+{2(c^2+4)\over 3s^2}\biggr)f_W\biggr\}
+\frac{3m_W^4}{8\pi^2\alpha\Lambda^2(m_H^2-m_W^2)}f_{\Phi,2}
\nonumber \\
R_{T3}&=& -{3\over 8 \pi c^2} {m_H^4\over\Lambda^2 (m_H^2-m_Z^2)}\biggl\{
f_B+{m_W^2\over m_H^2-m_W^2}f_W\biggr\}
+\frac{3m_H^4m_Z^2s^2}{8\pi^2\alpha\Lambda^2(m_H^2-m_Z^2)(m_H^2-m_W^2)}f_{\Phi,2}
\nonumber
\end{eqnarray}
\begin{flalign}
\quad \;\, R_{U1}=& \;\;
  \frac{m_Z^2}{24\pi\Lambda^2c^4}
  \left\{
    \left[ 48c^6f_{WW}-\rh^2\left( 2s^2c^4f_B+2\left( c^4+c^6\right)f_W+24c^6f_{WW}
      \right) 
      \right.\right.\nonumber\\&\left.\left.
      -\rh^4\left( s^2c^4f_B+\left( c^4+c^6 \right)f_W \right)\right]
      \rh\sqrt{4-\rh^2}\cos^{-1}\left( \frac{m_H}{2 m_Z} \right)
  \right.
  \nonumber\\&
  \left.
  +4\left[ -24c^4f_{WW}+2\rh^2c^2\left( f_W+6f_{WW} \right)+\rh^4f_W \right]
  \right.\nonumber\\&\left.\times
  \rh\sqrt{4c^2-\rh^2}\cos^{-1}\left( \frac{m_H}{2m_W} \right)
  \right.
  \nonumber\\&
  \left.
  -2\left[ \left( 80c^8+116c^6+90c^4+22c^2-11 \right)f_W+216f_{WWW}g^2 \right]
  \sqrt{4c^2-1}\sin^{-1}\left( \frac{1}{2c} \right)
  \right.
  \nonumber\\&
  \left.
  +2\left[ \left( -40c^4+6c^2+1 \right)\frac{s^2}{c^2}f_B+36c^2\left( 4c^2-1
    \right)f_{WWW}g^2 \right]\sqrt{4c^2-1}\cos^{-1}\left( \frac{1}{2c} \right)
  \right.
  \nonumber\\&
  \left.
  +\left[ \frac{2}{3}c^2\left( 240c^6-121c^4-62c^2+33 \right)
    +11\pi\sqrt{4c^2-1}\left( 8c^4+2c^2-1 \right)\right]f_W
  \right.
  \nonumber\\&
  \left.
  -\left( 84c^6-95c^4+9c^2+2 \right)f_B+96s^2c^4f_{WWW}g^2
  \right.
  \nonumber\\&
  \left.
  +3\rh^2s^2c^4\left( f_B-f_W-16f_{WW} \right)
  +2\rh^4s^2c^2\left[ c^2f_B-(c^2+2)f_W \right]
  \right\}
  \nonumber\\&
  +\frac{f_{\Phi,2}}{\Lambda^2}\frac{m_Z^2s^2}{24\pi^2\alpha c^4}\rh
  \left[ -9c^4s^2\rh+2c^2s^2(1+c^2)\rh^3\right.
    \nonumber\\&\left.
    +2c^6\left[ \rh^4-4\rh^2+12 \right]\sqrt{4-\rh^2}\cos^{-1}\left( \frac{m_H}{2m_Z}
    \right)
    \right.\nonumber\\&\left.
    -2\left[ 12c^4-4c^2\rh^2+\rh^4 \right]\sqrt{4c^2-\rh^2}\cos^{-1}
    \left(\frac{m_H}{2m_W}\right)
    \right]
  \nonumber \\
\quad \;\, R_{U2} =& \;\; \frac{1}{12 \pi  c^2 \Lambda^2 m_W^4} \left\{-24 c^2 f_{WW} m_W^2 \left(2 m_W^4-4 
m_H^2 m_W^2+m_H^4\right)+36 \left(1-6 c^2\right) \
f_{WWW} g^2 m_Z^2 m_W^4
  \right. 
  \nonumber\\ &
  \left.
+\left(34 c^8-118 c^6+58 c^4-3 c^2-1\right) f_B m_Z^6
  \right. 
  \nonumber\\ &
  \left.
+\frac{c^2 f_W}{s^2 \left(m_W^2-m_H^2\right)} \left(2 m_H^2 s^2 \left(-3 \
m_W^6-6 m_H^2 m_W^4-m_H^4 m_W^2+m_H^6\right)
  \right. \right. 
  \nonumber\\ &
  \left. \left.
+\left(18 c^8-146 c^6+24 c^4+55 c^2-11\right) \
m_Z^6 (m_H^2-m_W^2)\right) \right\} \nonumber \\
&+\frac{m_Z^4s^2}{12\pi^2\alpha\Lambda^2 c^4(m_H^2-m_W^2)}
\left[ 2c^8-38c^6\rh^2+24c^4\rh^4-7c^2\rh^6
  \right.\nonumber\\&\left.+\rh^8 \right]f_{\Phi,2}
\nonumber\\
\quad \;\, R_{U3} =& \;\; \frac{m_H^2 s^2}{12 \pi \Lambda^2}
\left\{f_{WW}\frac{24 \
\left((c^2 + 1) m_H^2-4 m_W^2\right)}{m_W^2} 
+f_W \left[ \frac{18 m_Z^2 m_W^2}{(m_H^2-m_Z^2) (m_H^2-m_W^2)}
  \right. \right.
  \nonumber\\ &
  \left. \left.
+\frac{\left(c^4+2 c^2+2\right) m_H^4}{m_W^4}-6\right] +f_B \frac{
m_H^6-m_H^4 m_Z^2-6 m_H^2 m_Z^4-12 m_Z^6}{m_Z^6-m_H^2 m_Z^4}\right\} 
\nonumber\\
&+\frac{m_Z^6s^4}{12\pi^2\alpha\Lambda^2c^4(m_H^2-m_W^2)(m_H^2-m_Z^2)}
\rh^2\left[
-(c^4+c^2+1)\rh^8\right.\nonumber\\&\left.
+(c^6+8c^4+8c^2+1)\rh^6-c^2(7c^4+31c^2+7)\rh^4
\right.\nonumber\\&\left.+24c^4(c^2+1)\rh^2-36c^6
  \right]f_{\Phi,2}
\end{flalign}

\newpage
\bibliographystyle{unsrt}
\bibliography{paper_eff}

\end{document}